# Alternative origins of polarity in compressively strained $SrTiO_3$-$RENiO_3$ capacitors

Evgenios Stylianidis[1,2], Panagiotis Koutsogiannis[3,4], Alexander Lione[5], Felix Risch[6], Laura Hechler[1,2], Igor Stolichnov[6], Jorge Íñiguez-González[7,8], José A. Pardo[3,9], Nicholas C. Bristowe[5], César Magén[3,4], Pavlo Zubko[1,2]

[1] Department of Physics and Astronomy, University College London, Gower Street, London WC1E 6BT, United Kingdom
[2] London Centre for Nanotechnology, 17-19 Gordon Street, London WC1H 0AH, United Kingdom
[3] Instituto de Nanociencia y Materiales de Aragón (INMA), CSIC-Universidad de Zaragoza, 50009 Zaragoza, Spain
[4] Departamento de Física de la Materia Condensada, Universidad de Zaragoza, 50018 Zaragoza, Spain
[5] Condensed Matter Physics, Physics Department, Durham University, Durham, United Kingdom
[6] Nanoelectronic Devices Laboratory (NanoLab), Ecole Polytechnique Fédérale de Lausanne (EPFL), 1015, Lausanne, Switzerland
[7] Smart Materials Unit, Luxembourg Institute of Science and Technology (LIST), Avenue des Hauts-Fourneaux 5, L4362, Esch-sur-Alzette, Luxembourg
[8] Department of Physics and Materials Science, University of Luxembourg, 41 Rue du Brill, L-4422 Belvaux, Luxembourg
[9] Departamento de Ciencia y Tecnologia de Materiales y Fluidos, Universidad de Zaragoza, 50018 Zaragoza, Spain

## Abstract

Since its original prediction 25 years ago, room-temperature out-of-plane ferroelectricity in compressively strained $SrTiO_3$ remains an ongoing pursuit. In this work, we investigate the structural, electrical and electromechanical properties of highly strained epitaxial $SrTiO_3$ capacitors with rare earth nickelate electrodes. The $SrTiO_3$ layers experience compressive strains up to −3% and exhibit pronounced tetragonality, comparable to that of bulk $PbTiO_3$. Variable-temperature electrical measurements and room-temperature piezoresponse force microscopy reveal butterfly-shaped capacitance–voltage hysteresis and domain-like electromechanical response typical of ferroelectric materials. However, the overall behavior is inconsistent with a stable ferroelectric state. We therefore propose an alternative mechanism for the observed polarity in our samples based on spatially inhomogeneous internal fields. Our first-principles calculations show that such fields may arise from charge discontinuities between the formally charged $NdNiO_3$ layers and charge-neutral $SrTiO_3$ layers.

## *Introduction*

The theoretical prediction of strain-induced ferroelectricity in $SrTiO_3$ (STO) by Pertsev et al. [1], and its subsequent observation in epitaxial films under tensile strain a few years later [2], is a striking example of the great potential of strain engineering for designing, tuning and inducing new materials properties and functionalities [3]. At room temperature, bulk STO has a cubic perovskite structure and a dielectric permittivity of around 300. At 105 K, it undergoes a ferroelastic transition to a tetragonal state driven by cooperative rotations of the oxygen octahedra [4–6]. On further cooling, its dielectric permittivity increases dramatically as the material approaches a paraelectric-to-ferroelectric phase transition [7], which, however, is suppressed by quantum fluctuations [8,9]. Ferroelectricity in STO can be stabilized by small perturbations, such as minute isovalent substitution on the Sr site (e.g. with Ca) [10] or O-18 isotope exchange [11]. Furthermore, even slight off-stoichiometry (e.g. due to Sr vacancies) can locally lower the symmetry and form polar nanoclusters that percolate into a macroscopic polar phase [12–14].

In 2000, Pertsev et al. proposed that ferroelectricity in STO could also be stabilised by epitaxial strain [1,15]. Using phenomenological theory, they predicted a complex phase diagram for biaxially-strained STO with several ferroelectric phases that exhibit in-plane polarization components under tension and out-of-plane polarization under compression. Soon after, predictions of strain-induced ferroelectricity were corroborated by first-principles calculations [16,17], and near room-temperature ferroelectricity was demonstrated experimentally by Haeni et al. in STO films on $DyScO_3$ (around +1% tensile strain) [2]. Since then, several aspects of the in-plane ferroelectric phase in tensile-strained STO have been explored, including its relaxor-like nature [18] and its potential for electrocaloric cooling [19].

As one of the few oxides that can be grown epitaxially directly on silicon [20,21], strain-induced ferroelectricity in STO is particularly attractive for applications in electronics. Such applications typically require an out-of-plane polarisation at room temperature that should be achievable with a sufficiently large compressive strain, and indeed ferroelectricity has been explored in STO compressively strained on silicon [21]. However, to date, a convincing demonstration of a macroscopically switchable, room-temperature strain-induced out-of-plane polarisation remains elusive, and there is generally less consensus in the literature on the behaviour of compressively strained STO. Polarisation-voltage hysteresis has been reported at cryogenic temperatures in a 60 nm thick STO film grown on $NdGaO_3$, which imparts around -1% compressive strain [22]. The appearance of polarisation switching in this case was accompanied by a broad, relaxor-like dielectric anomaly. Optical studies on the same system found no evidence of a ferroelectric transition down to 5 K [23], contrary to conclusions of a combined XRD and Raman study [24]. Isolating the effect of strain is particularly challenging because ferroelectric-like behaviour is easily induced by defects, even in nominally unstrained films and particularly at reduced dimensions [14,22,25]. For compressively strained films, both enhancement and suppression of polarity due to non-stoichiometry have been reported [26,27]. Additionally, because large compressive strains can only be sustained in sufficiently thin films, electrical measurements on macroscopic capacitors are frequently hindered by leakage. Polarisation switching is therefore usually demonstrated using scanning probe techniques performed on structures with an exposed top surface, which results in poorly controlled electrostatic boundary conditions that further complicate the interpretation [28]. It is therefore not yet clear whether a stable ferroelectric state can be induced by compressive strain in a capacitor structure at room temperature, and the typical questions of polarisation stability and domain formation under imperfect screening, which have been extensively studied in conventional ferroelectrics, remain unexplored.

In this work, we study the properties of compressively strained epitaxial STO capacitors with rare earth nickelate ($RENiO_3$) electrodes. Our STO layers are subjected to compressive strains as large as −3% and display a unit cell tetragonality comparable to that of bulk $PbTiO_3$. We investigate in detail the electrical and electromechanical response of our parallel-plate capacitor structures using a combination of variable-temperature electrical measurements and room-temperature piezoresponse force microscopy (PFM). While our samples exhibit several features of typical ferroelectrics, including an electromechanical response characteristic of polydomain ferroelectric films and butterfly-shaped capacitance-voltage hysteresis at room temperature, the overall temperature-dependent behaviour is not consistent with ferroelectricity. Using first-principles calculations, we examine the effects of the interfacial charge discontinuity between the formally charged layers of $NdNiO_3$ and the charge neutral layers of $SrTiO_3$ and propose an alternative origin for our experimental observations.

## ***Experimental Results***

### ***Growth and structural characterization***

$RENiO_3$/$SrTiO_3$/$RENiO_3$ heterostructures were grown using off-axis radiofrequency magnetron sputtering on (001)-oriented $(La_{0.18}Sr_{0.82})(Al_{0.59}Ta_{0.41})O_3$ (LSAT), (110) $NdGaO_3$ (NGO) and $(001)_{pc}$ $LaAlO_3$ (LAO) substrates with room-temperature pseudo-cubic (pc) lattice parameters of 3.868 Å, 3.863 Å and 3.7896 Å, respectively. The in-plane biaxial strain imposed on STO ($a = 3.905$ Å in bulk) due to the lattice mismatch with these substrates is $-0.9\%$, $-1.1\%$ and $-2.9\%$ on LSAT, NGO and LAO, respectively. Perovskites $LaNiO_3$ (LNO) and $NdNiO_3$ (NNO), both members of the wider family of the perovskite rare earth nickelates ($RENiO_3$), were chosen as metallic layers. To achieve the best lattice parameter matching, the NNO layers ($a_{pc} = 3.81$ Å) were chosen for the heterostructures grown on LAO, while LNO layers ($a_{pc} = 3.84$ Å) were used for those on NGO and LSAT.

Except for LNO, which is metallic at all temperatures, perovskite rare-earth nickelates are characterized by a thermally driven metal-insulator transition (MIT) [29,30]. Although not the subject of the present study, combining $RENiO_3$ electrodes such as NNO with ferroelectric materials presents a fascinating opportunity to explore the effect of the MIT on screening of the ferroelectric polarization. For the purpose of this work, however, we grew NNO layers with an almost fully suppressed MIT, engineered by varying the growth temperature (Supplementary Figure B1). This engineered fully-metallic NNO layer is a suitable electrode material for our studies. Parenthetically, the effect of the metal-insulator transition of NNO electrodes on impedance measurements of capacitors grown on LSAT is discussed in the Supplementary Section C.

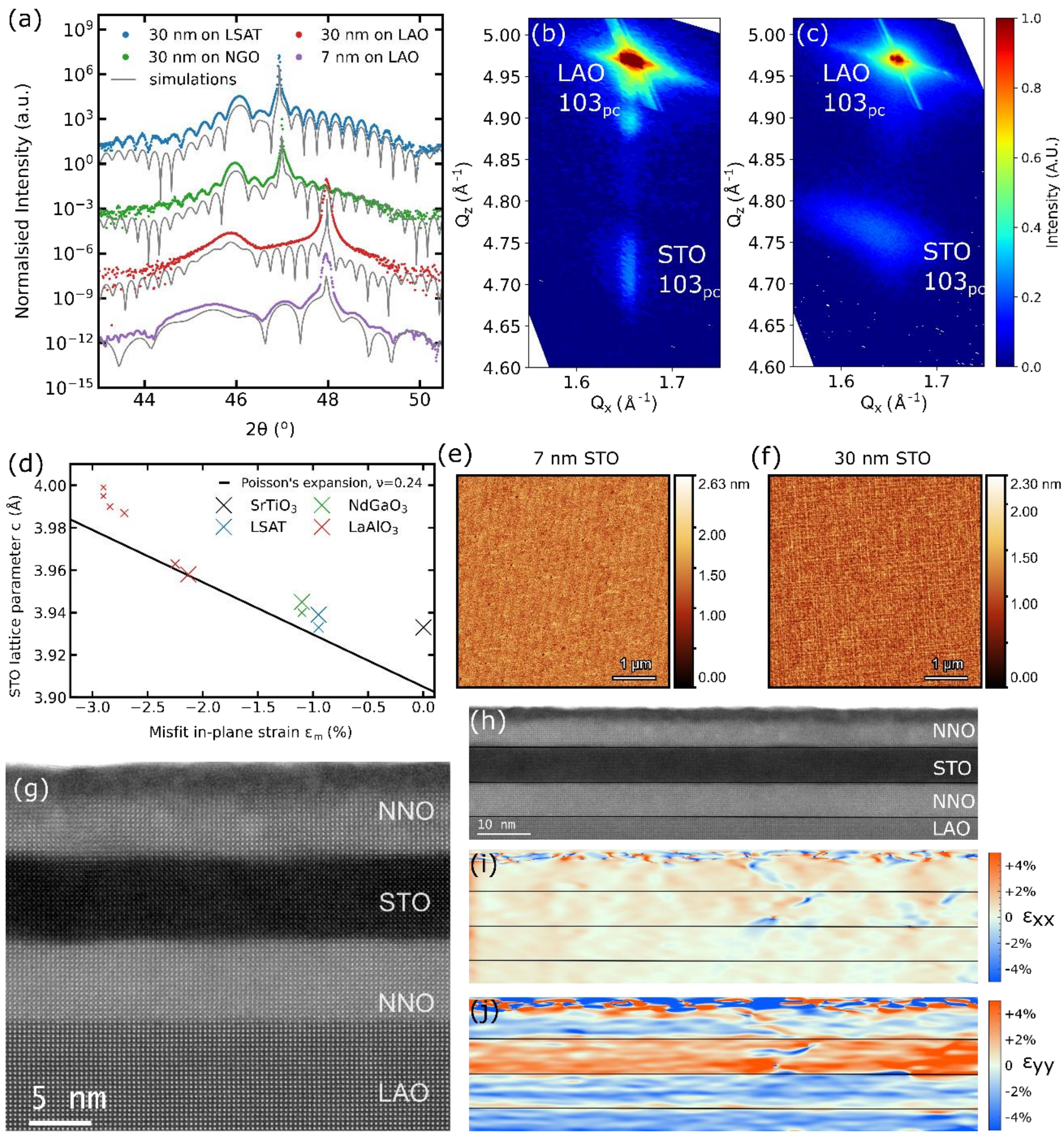

**Figure 1:** Structural characterization of $RENiO_3/SrTiO_3/RENiO_3$ heterostructures. (a) X-ray diffraction θ-2θ scans around the 002 substrate reflection and simulations performed using a kinematical diffraction model [31]. The legend details the thickness of the $SrTiO_3$ layer and the substrate. The $RENiO_3$ layers are 6 nm thick. (b) and (c) Reciprocal space maps around the $103_{pc}$ LAO reflection for $NdNiO_3/SrTiO_3/NdNiO_3$ heterostructures on LAO with 7 nm and 30 nm-thick $SrTiO_3$, respectively. (d) STO out-of-plane lattice parameter as a function of the in-plane strain determined experimentally (×) and expected (black solid line) from linear elasticity theory assuming a Poisson ratio $\nu = 0.24$. The size of the markers is proportional to the thickness of each $SrTiO_3$ film. (e) and (f) Atomic force microscope images for the samples in (b) and (c), respectively. (g) and (h) Cross-sectional HAADF STEM images of the $NdNiO_3/SrTiO_3/NdNiO_3$ heterostructure with 7 nm $SrTiO_3$ grown on LAO. (i) and (j) In-plane $\varepsilon_{xx}$ and out-of-plane $\varepsilon_{yy}$ deformation maps, respectively, generated from the HAADF-STEM image in (h) using geometrical phase analysis (GPA).

A series of heterostructures were grown with constant LNO and NNO thicknesses of around 6 nm, and varying STO thickness, ranging from 7 nm to 30 nm. The thickness of each layer was determined using X-ray reflectivity (Supplementary Figure D1), while X-ray diffraction was used to assess the structural quality. Figure 1 (a) shows diffractograms around the $002_{pc}$ substrate reflections for representative samples. Samples on NGO and LSAT with 30 nm STO exhibit clear Laue oscillations, indicative of coherent crystalline heterostructures. On LAO, the structural quality depends on the thickness of STO (Supplementary Figure D2): while the sample with 7 nm STO shows Laue oscillations, with increasing STO thickness the oscillations fade out and a broader film peak is observed, characteristic of structural relaxation. Rocking curves around the 002 peak of the STO films show enhanced diffuse intensity for thicker STO, reflecting the disruption of the long-range coherence (Supplementary Figure D3). Reciprocal space maps around the off-specular $103_{pc}$ reflections (Figure 1 (b), (c) and Supplementary Figure D4) further reveal that the sample with 7 nm STO is essentially fully strained on the substrate, whereas 30 nm-thick STO undergoes significant relaxation.

Atomic force microscopy (AFM) (Figure 1 (e-f)) reveals stripe-like features on the surface of the 7 nm STO heterostructure, likely related to the substrate miscut steps and terraces, while a cross-hatch surface morphology is observed for the sample with 30 nm STO, reminiscent of the dense network of misfit dislocations often observed in relaxed films [32,33]. We further employed scanning transmission electron microscopy (STEM) to examine the atomic structure of the samples. As shown in Figure 1 (g) and (h), the sample with 7 nm STO on LAO exhibits sharp interfaces and good overall structural coherence, with only sparse misfit dislocations (appearing as local anomalies in the deformation maps in panels (i-j)). By contrast, the 30 nm STO sample shows dense clustering of misfit dislocations at the bottom NNO/STO interface (Supplementary Figure D6), consistent with the structural relaxation inferred from X-ray measurements and surface features observed with AFM.

We extracted the out-of-plane lattice parameters $c_{film}$ for all STO films by simulating the diffractograms using a kinematical diffraction model [31] (Figure 1 (a)). In-plane lattice parameters $a_{film}$ of the relaxed films were estimated after projecting the $103_{pc}$ film peak intensity along $Q_x$, and the effective in-plane strain was calculated using $\varepsilon_{film} = a_{film}/a_{bulk} - 1$, where $a_{bulk} = 3.905$ Å is the bulk lattice parameter of STO. A very large tetragonality ($c_{film}/a_{film}$ ratio) of 1.055 is observed for the 7 nm STO film grown on LAO. Notably, this value is comparable to that of $PbTiO_3$ which has a bulk tetragonality of 1.06 at room temperature in the ferroelectric phase [34]. Figure 1 (d) compares the measured out-of-plane lattice parameters as a function of biaxial strain with those predicted by elasticity theory using a Poisson ratio of $\nu = 0.24$. Evidently, our strained STO films have enhanced out-of-plane lattice parameters compared to these theoretical estimates.

***<u>Macroscopic electrical properties:</u>***

To investigate the ferroelectric properties of our compressively-strained STO films, we perform a series of macroscopic electrical measurements on the capacitor structures. To improve the electrical contact to the top electrode, we coat the samples with a thin (~10 nm) Pt layer prior to defining the circular top electrodes via Ar-ion milling (see inset of Figure 2 (b) and Methods). Figure 2 (a) and (b) shows a plot of the effective dielectric permittivity ε calculated directly from the measured capacitance (see Methods), and the loss tangent for capacitors with 30 nm STO thin films grown on LSAT, NGO and LAO, and a 7 nm STO film grown on LAO. Irrespective of the substrate, the behavior of all 30 nm films is qualitatively similar, with a broad permittivity maximum around 50 K suggestive of a possible onset of a ferroelectric phase transition below this temperature. However, no evidence of ferroelectric-like hysteresis was observed in capacitance-voltage measurements down to 11 K (Supplementary Figure E1).

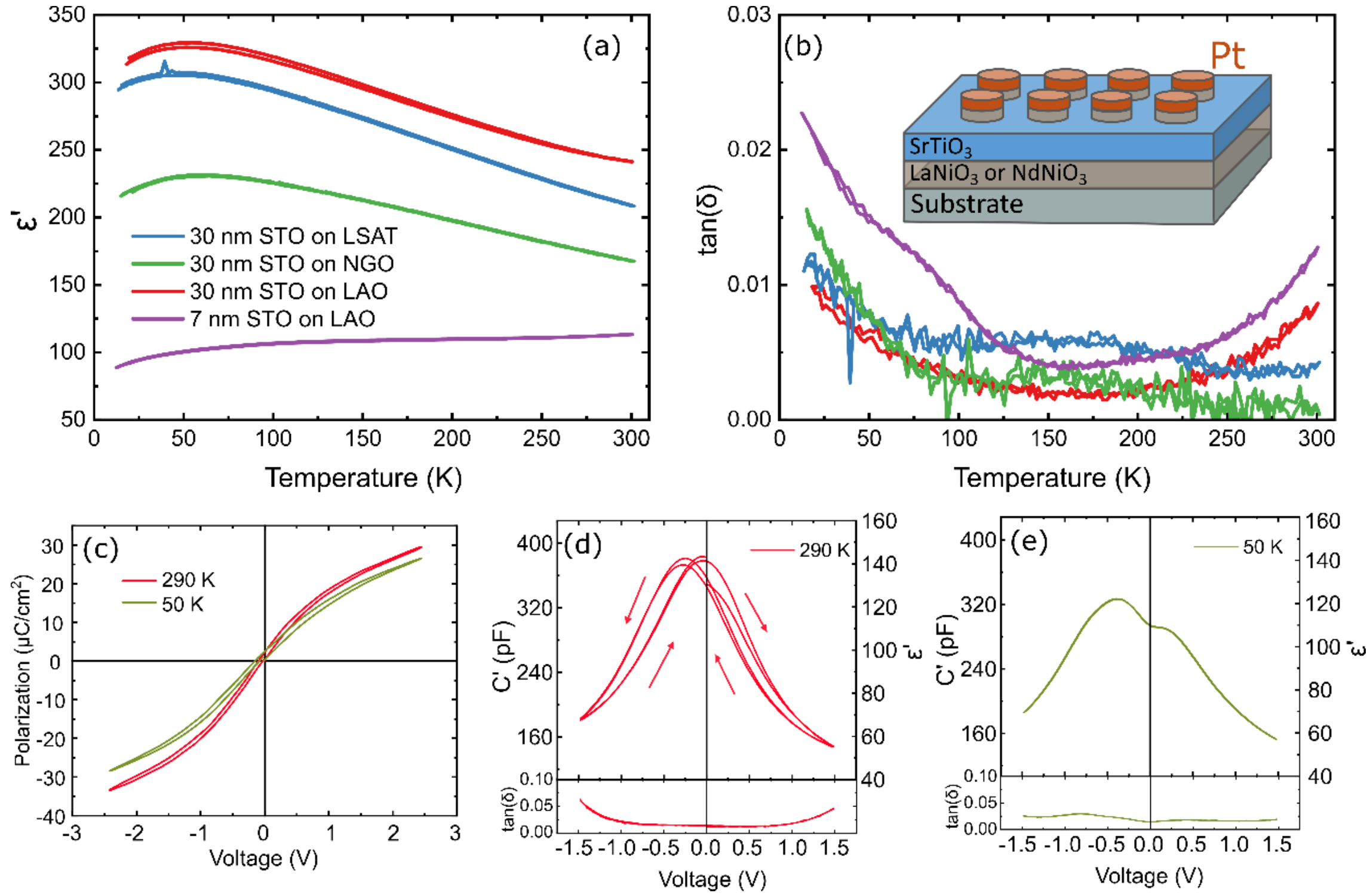

**Figure 2:** (a) Relative permittivity and (b) loss tangent as a function of temperature measured with a 10 mV and 1 kHz excitation bias for different heterostructures. The inset in (b) shows schematically the fabricated circular parallel-plate capacitors. (c) Polarization-voltage measurements at 290 K and 50 K measured with a 1 kHz triangular voltage pulse and (d), (e) capacitance-voltage measurements at 290 K and 50 K, respectively, for a Pt-coated $NdNiO_3/SrTiO_3$(7 nm)/$NdNiO_3$ heterostructure grown on $LaAlO_3$.

By contrast, the capacitor with 7 nm-thick STO grown on LAO behaves differently. The effective permittivity increases slightly with increasing temperature, which suggests a possible permittivity maximum, the hallmark of a ferroelectric transition, above room temperature. However, we highlight the small permittivity variation across the explored temperature range, which likely reflects significant contribution from interfacial capacitance effects with unknown temperature dependence. High-temperature measurements (not shown here) reveal a further slight increase in capacitance up to 400 K. For even higher temperatures, repeated measurements on different samples gave different results with appearance of additional relaxations, akin to space charge effects observed in bulk perovskites at elevated temperatures [35], or due to enhanced DC conductivity across the STO layers.

Polarisation-voltage (P-V) measurements for a 7-nm thick STO capacitor on LAO are shown in Figure 2 (c). Despite the small STO thickness, no discernible contribution from conductivity is observed in these 1 kHz measurements, indicating highly insulating behaviour. At 290 K, a slim P-V loop with negligible hysteresis and remanence is observed. Such behaviour can be characteristic of non-linear dielectrics with no spontaneous polarisation, relaxors, or ultrathin ferroelectric capacitors where due to imperfect screening the macroscopic polar state is unstable to the formation of nanoscale domains. However, in the case of multidomain ferroelectrics, typically, hysteresis eventually appears at lower temperatures [36], which is not the case for our STO films, as evident from the 50 K measurement.

Capacitance-voltage measurements on the same sample at 290 K and 50 K are shown in Figure 2 (d) and (e), respectively. Surprisingly, here, butterfly-shaped hysteresis is observed at 290 K, reminiscent of the typical C-V curves found in ferroelectric capacitors. Upon cooling, however, the hysteresis shrinks and a fully reversible behaviour is observed at 50 K, albeit with an asymmetric bimodal shape.

***<u>Piezoresponse force microscopy</u>***

To probe the local polarity of our films and the possibility of domain formation, we perform piezoresponse force microscopy (PFM) measurements on both asymmetric structures (with a bottom electrode and exposed STO surface) and symmetric capacitor structures. The results of PFM measurements on a sample with an exposed surface are detailed in Supplementary Figure F1. The as-grown sample exhibits a uniform piezoresponse. Upon application of $\pm$8 V while scanning the tip, the sample can be poled. Oppositely poled regions exhibit similar amplitude, 180° phase contrast and amplitude minima at their boundaries, as is typical of written ferroelectric domains. However, after 1.5 hours, the phase contrast fades and the amplitude decreases, indicating possible charging effects.

The results of measurements on capacitor structures are shown in Figure 3 for a sample with NNO electrodes and in Supplementary Figure F2 for a sample with LNO electrodes. The Pt-coated capacitors were connected to an external wire, which is held at the same potential as the conducting AFM tip, as described in Methods and sketched in Figure 3 (b). The parallel-plate capacitor geometry ensures that the applied field is homogeneous across the STO and eliminates the ill-defined electrostatic boundary conditions found at bare surfaces [37,38]. Although the PFM resolution is inherently more limited in a capacitor geometry, this technique has proven highly effective for probing the piezoresponse of complex domain patterns under real device conditions [37,39].

Vertical PFM images acquired at room temperature over the Pt/NNO electrode for a capacitor with a 7 nm STO film grown on LAO are displayed in Figure 3 (a). During the measurements, the 1.2 V AC excitation bias was superposed on the DC bias. Already at 0 V DC bias, clear domain-like PFM phase contrast is observed with reduced PFM amplitude at the boundaries. Similar behaviour has been reported in ferroelectric $PbTiO_3$ films with “up” and “down” polarized domains, induced by imperfect screening [40]. Application of negative DC bias to the top NNO electrode increases the proportion of one phase state over the other until a homogeneous “up” state is reached. Reducing the magnitude of the bias again causes the mixed phase state to reappear, and subsequent application of positive bias gradually results in a homogeneous “down” state. By thresholding the phase images, we extracted the fraction of one phase as a function of the DC bias, as depicted in Figure 3 (b). The lateral shift of the switching curve towards negative DC bias suggests the presence of a local positive built-in bias. Interestingly, after removing the +1 V DC bias, back-switching is observed, and the capacitor shows a majority "up" state (blue triangle at 0 V DC bias), suggesting alteration of the imprint across the capacitor.

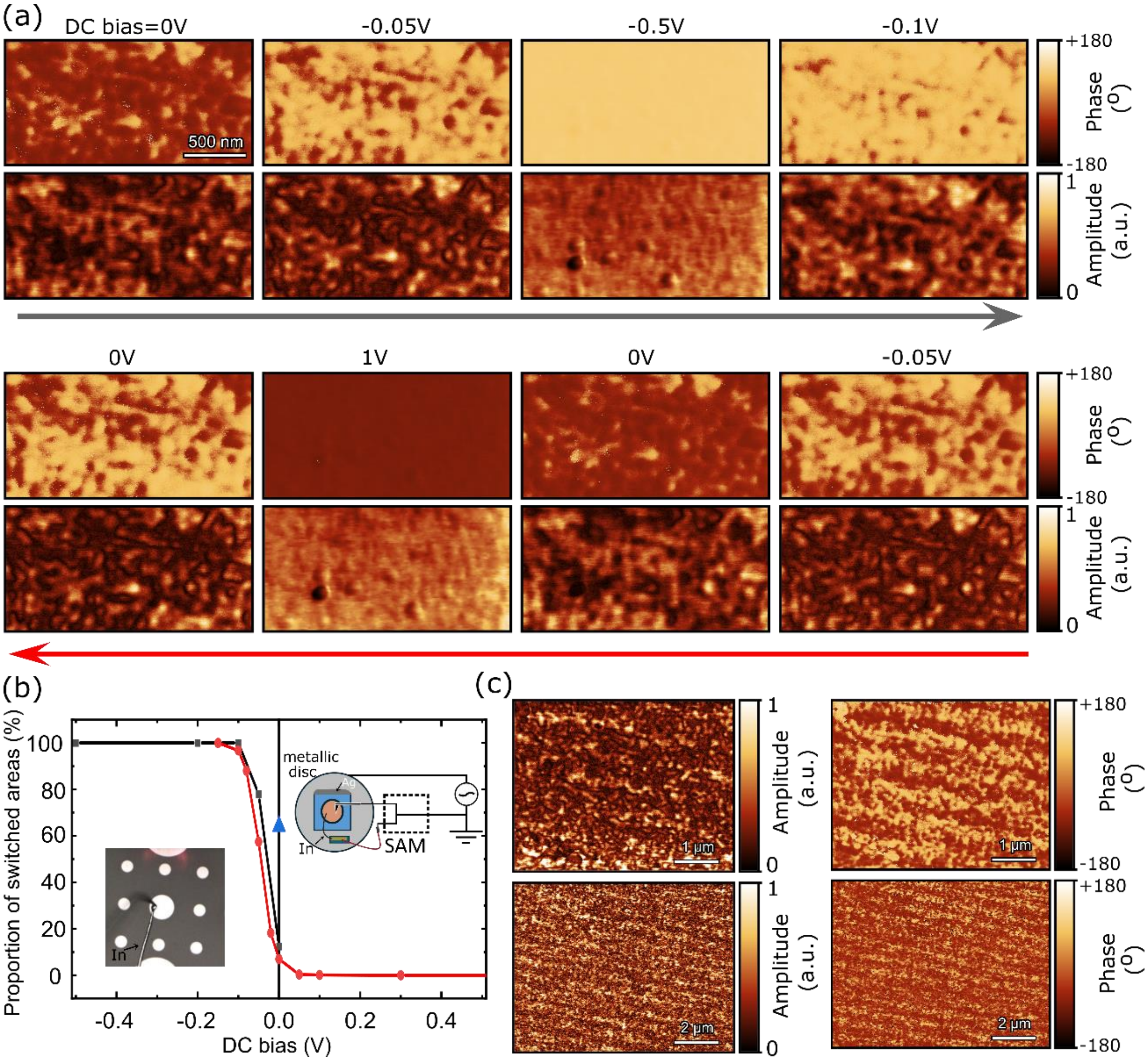

**Figure 3:** (a) Piezoresponse force microscopy measurements performed on a Pt-coated $NdNiO_3/SrTiO_3$(7 nm)/$NdNiO_3$ heterostructure grown on $LaAlO_3$, measured with 1.2 V and 297 kHz AC bias. (b) Proportion of switched areas as a function of the applied DC bias. Bottom left inset: optical image of the top surface of the sample. Top right inset: schematic of the experimental configuration of our PFM experiments on parallel-plate capacitors. (c) Larger scale PFM images measured at 0 V DC bias for the same sample as in (a).

Upon departing from a monodomain state, we found that domains always re-nucleate at the same locations, just like a heterogeneous switching process often found in traditional ferroelectric films [41,42]. Moreover, larger-scale images recorded at 0 V DC bias, Figure 3 (c), revealed a long-range variation of the observed electromechanical response. The domains appear to form a stripe-like modulation of the majority and minority states: in one stripe, the "up" domains dominate forming a matrix with bubble-like "down" domains in it, while in the adjacent stripe, the "down" domains become the matrix and the "up" domains become the bubbles. We note that the surface of the same sample before the deposition of Pt displayed a stripe-like morphology (Figure 1 (g)), which was attributed to miscut steps and terraces of the substrate. However, due to the presence of Pt we were unable to determine the surface topography of the epitaxial stack at the exact location the PFM was performed, and therefore we have not systematically investigated the correlation between topography and electromechanical response.

PFM experiments were also performed for heterostructures grown on LAO with STO thicknesses up to 30 nm. All samples showed similar electromechanical behaviour, independent of STO thickness or strain state. In the 30 nm STO sample, the response (not shown here) resembled the cross-hatch surface morphology (Figure 1 (g)), which we attributed earlier to the dense network of misfit dislocations. Our PFM studies show that the STO films grown on LAO, under a compressive strain that ranges from -3% to -2.1%, exhibit a similar electromechanical response. To test whether this arises from room-temperature ferroelectricity in STO, we performed similar PFM experiments on STO capacitors grown on NGO and LSAT. Both theory [1] and experiments [22,27] suggest that STO should remain paraelectric at room temperature at these strains. Surprisingly, we observed a very similar electromechanical response (Supplementary Figure F3).

## ***Discussion***

According to the calculated phase diagram of Pertsev et al. [1], STO under a compressive strain in excess of -1.8% should display a paraelectric-to-ferroelectric transition above room temperature. Therefore, our fully-strained 7-nm thick STO samples are expected to be ferroelectric already at room temperature. At first sight, these capacitors show several common signatures of room-temperature ferroelectricity, including highly enhanced tetragonality, clear domain-like features in PFM, and butterfly-shaped C-V hysteresis curves. However, a closer examination reveals some unusual behavior at odds with conventional ferroelectricity. First, the suppression of the C-V hysteresis upon cooling and the absence of P-V hysteresis at all temperatures are inconsistent with a switchable spontaneous polarisation. Instead, the room-temperature hysteretic C-V suggests a thermally-activated process. The absence of P-V hysteresis further supports this view: the P-V loop, acquired with a 1 kHz triangular voltage pulse, is completed within 1 ms, while the C–V curve, measured using an AC bias atop a slowly incremented DC bias, takes ~1 min. Such long timescales enable slower processes, such as electronic or ionic charge redistribution, able to produce spurious hysteresis in C–V measurements [28]. As will be discussed in more detail elsewhere [43], we speculate that the observed C-V hysteresis is related to charge injection that leads to dynamic voltage offsets during field cycling, as previously reported for $(Ba,Sr)TiO_3$ capacitors [44].

Second, although both strain-induced polarization and antiferrodistortive ordering (associated with oxygen octahedral tilts typical of bulk STO) are expected to lead to an increase in tetragonality (see discussion in Supplementary Section G), tetragonality enhancements in STO are also frequently reported to be associated with cation non-stoichiometry and/or presence of oxygen vacancies [45,46]. There is some disagreement as to the effect that such defects have on polarity and ferroelectricity in STO, with some studies reporting that only near-stoichiometric samples display strain-induced ferroelectricity [27], while others propose that defects can enhance the ferroelectric properties [22,25,26,47]. The lattice parameters of our fully strained films on STO are almost identical to those reported recently in similar films by Chen et al. and attributed to defect-enhanced ferroelectricity [26]. However, our macroscopic electrical measurements clearly show that our samples are not ferroelectric down to 15 K.

Finally, our room-temperature PFM measurements on fully strained STO capacitors grown on LAO exhibit all the textbook signatures of ferroelectric domains and their response to applied field. However, observation of similar behavior for films on NGO and LSAT substrates, which are not expected to provide sufficient compressive strain to induce room-temperature ferroelectricity, casts doubt on the ferroelectric origin of this signal.

We propose that the observed PFM contrast may arise from a spatial variation of local built-in fields. It is important to note that PFM does not distinguish between spontaneous and induced polarization. Thus, any spatially varying built-in field will induce a spatially varying polarization and associated piezoelectric response. If the local built-in field varies in sign across the sample, the piezoresponse will vary in phase and hence produce domain-like contrast, as illustrated schematically in Figure 4. When an external field is applied to the sample, the piezoresponse will be determined by the net field resulting from the superposition of the external and local built-in fields. Thus, a negative external field will increase the size of the regions with a net negative field and decrease the size of the regions with a net positive field, leading to apparent "domain growth" and motion of the boundaries where the response is zero (Supplementary Figure H1).

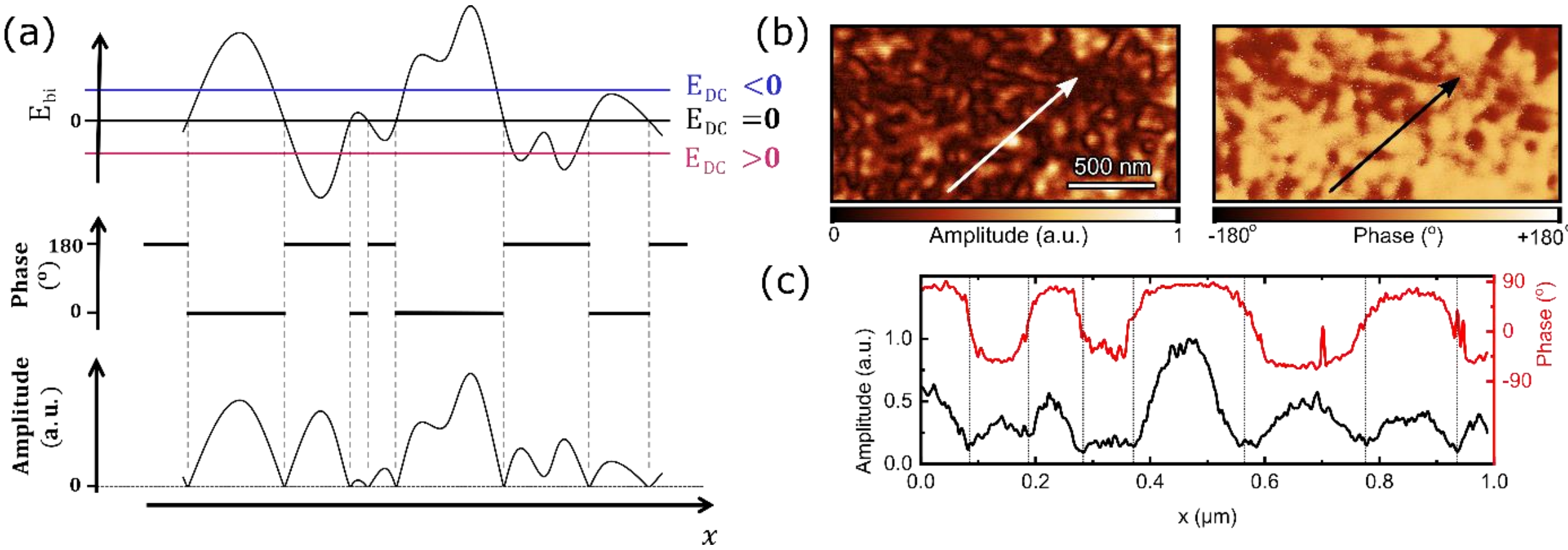

**Figure 4:** Electromechanical response model. (a) Schematic depiction of how a spatially varying built-in field can give rise to an electromechanical response reminiscent of a polydomain ferroelectric. The blue and red lines represent the shifted baseline upon application of an external field. The expected behaviour of the PFM phase and amplitude under applied bias is shown in Supplementary Figure H1. (b) Experimental PFM amplitude and phase images of our $SrTiO_3$ films measured atop the $NdNiO_3$/Pt top electrode with no applied external DC bias and a 1.2 V and 297 kHz AC bias. (c) Amplitude and phase profiles along the white and black lines shown in (b), respectively.

To explore the possible origin of such built-in fields and reasons for the absence of the predicted ferroelectricity in our samples, we turn to first-principles simulations.

## ***First principles simulations***

The phenomenological models used by Pertsev et al. [1] and others [48] to predict a strain-induced polar ground state in STO rely on a set of empirical parameters and assume a homogeneous system with ideal short-circuit boundary conditions, thus neglecting any effects associated with the metal-dielectric interfaces and defects. To simulate a more realistic system, we perform first-principles calculations on both pure STO and capacitor structures with an explicit simulation of the nickelate electrodes.

First, we re-examine the stability of the strain-induced polar phase by computing the ground state phase diagram of STO using first-principles methods (Figure 5 (a) and Supplementary Figure I1). We find that under −3% compressive strain, an out-of-plane polar phase is indeed the ground state at 0 K, consistent with previous works [16,17], but it is less than 8 meV per formula unit (p.f.u.) away from a competing antiferrodistortive non-polar phase. This small energy difference indicates that the polar state is relatively fragile and may be difficult to observe at finite temperatures. Parenthetically, under tensile strain, the polar ground state phase shows enhanced stability: the energy difference increases from around 11 meV p.f.u. at +1.3% tensile strain to 39 meV p.f.u. at +2.6% tensile strain. Of course, our calculations focus on ideal STO, ignoring the effects of defects or off-stoichiometry, inevitably present in real materials. The versatile perovskite structure can accommodate both cation and oxygen vacancies, whose formation energies decrease in highly-strained thin films [49]. Zhou and collaborators investigated by *ab initio* calculations the ground state phase diagram of biaxially strained oxygen deficient $SrTiO_{3-\delta}$ [50]. They found that, while oxygen deficiency has little effect in the tensile regime, out-of-plane ferroelectricity in the compressive regime is highly sensitive to oxygen vacancy content and position within the unit cell: for $SrTiO_{2.75}$ at 0 K, apical vacancies suppress ferroelectricity up to −4% strain, while equatorial vacancies suppress it up to −3%. The large enhancement in tetragonality of our films is consistent with non-stoichiometry that likely contributes to the absence of ferroelectricity in our samples.

Next, we investigate the effect of interfaces between the NNO electrodes and STO by simulating capacitor structures. A key feature of these interfaces is a change in the formal valences of the cations in NNO and STO and hence in the layer charges across it, as depicted in Figure 5 (f). While STO is composed of charge-neutral $SrO^0$ and $TiO_2^0$ layers, NNO can be modelled as consisting of alternating $NdO^+$ and $NiO_2^-$ layers with formal charges +1e and -1e, respectively, which are self-screened by the metallic free charges. In the extensively studied interface between insulating LAO and STO a similar change in formal layer charges leads to a polar discontinuity, which is compensated by a transfer of charge into the STO, making it conducting [51]. In metallic NNO, free carriers are already available to screen the formal layer charges. However, this screening is never perfect and the associated finite screening length leads to interfacial dipoles. Under short circuit boundary conditions, these dipoles result in an effective field across the STO layer and corresponding tilting of its bands. First-principles calculations by Geisler et al. for the similar $LaNiO_3$-$SrTiO_3$ system indeed confirmed that STO remains insulating for all types of interfaces, and that a band-offset of around 0.5 eV is formed across a stoichiometric $LaNiO_3$/$SrTiO_3$/$LaNiO_3$ trilayer with dissimilar polar interfaces [52].

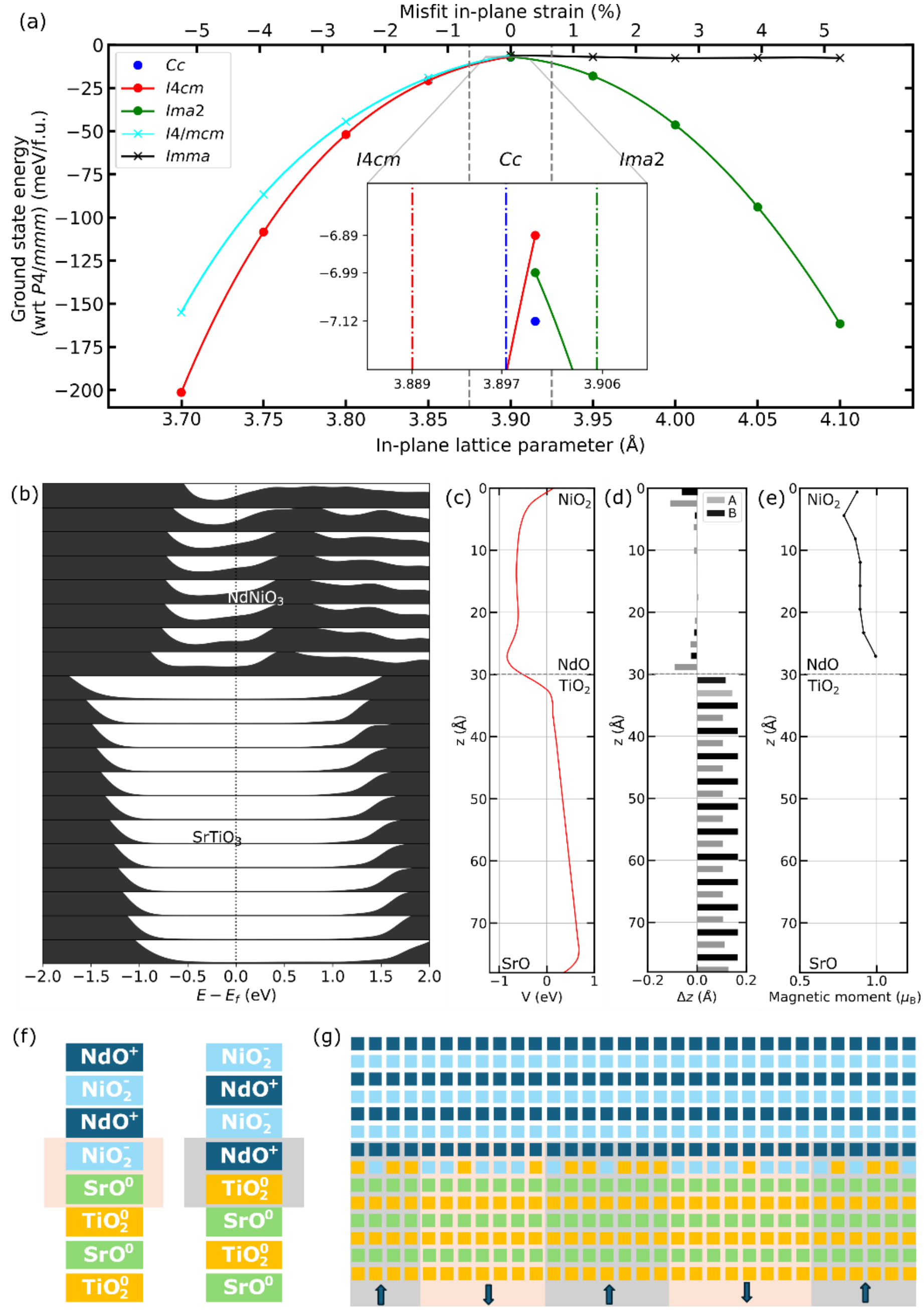

**Figure 5:** First-principles calculations. (a) Ground state strain phase diagram of $SrTiO_3$. Inset shows subtle energy differences (also in meV/f.u.) between phases at 3.90 Å, where vertical dashed lines indicate the bulk lattice parameters of each phase. The characteristics of each phase are shown in Supplementary Table I1. (b-e) Electronic and structural characteristics of a stoichiometric $NdNiO_3$-$SrTiO_3$ system as a function of z position (direction along $\hat{c}_{001}$): (b) layer projected density of states (PDOS) near $E_F$, where each layer has chemical formula $2ABO_3$, and the y-axis of each layer plot ranges from 0 to 3 states/eV, (c) macroscopically averaged potential (averaged over mean $SrTiO_3$ interlayer distance (= 4.0 Å)), (d) A- and B-site polar displacements, and (e) layer magnetic moments along $\hat{c}_{001}$. (f) The two possible $SrTiO_3$-$NdNiO_3$ heterointerfaces. (g) Schematic illustration of the proposed mixed-terminated $SrTiO_3$-$NdNiO_3$ heterostructure, capable of inducing a mixed polarization profile.

To examine the effects of the interfaces in our $NdNiO_3/SrTiO_3/NdNiO_3$ capacitors we conducted first-principles calculations on strained $NdNiO_3/SrTiO_3$ supercells, with in-plane lattice constants fixed to those of LAO and short-circuit periodic boundary conditions. We examined two model systems: a "non-stoichiometric" structure with identical top and bottom interfaces, either SrO-$NiO_2$ or $TiO_2$-NdO, and a "stoichiometric" one with dissimilar interfaces. NNO was treated as a ferromagnetic metal in all systems (see methods). In the non-stoichiometric cases, a net inversion symmetry is preserved, forcing a zero net polarization in the STO. Our computations revealed a large STO band offset between each interface termination, equal to approximately 1.59 eV, and non-uniform charge distribution within the NNO layers (Supplementary Figures I2 and I3). This indicates that despite the metallic character of NNO, the layer charge discontinuity between NNO and STO is not fully screened, and produces a similar effect to the polar discontinuity between STO and LAO [51]. A detailed discussion can be found in Supplementary Section I.

The results for the stoichiometric case are shown in Figure 5 (b-e). Here, the interfacial dipoles generate an internal field in STO, producing a band offset of 0.69 eV. In this case, the internal field strongly polarizes the STO slab, pinning the polarization from the NdO-$TiO_2$ to the SrO-$NiO_2$ interface. We note that the band offset of 0.69 eV of the stoichiometric case is significantly lower than the 1.59 eV estimated between the non-stoichiometric systems, a result of the screening caused by the polarization in STO. We found that even if the polarization is initialized in the opposite direction, it automatically relaxes to the orientation imposed by the internal field. Notably, the induced polarization in STO is amplified compared to the predicted polar ground state arising from strain effects alone (see Supplementary Table I2). We note that our DFT-predicted tetragonality enhancement, arising from the combination of polar and AFD modes and induced polarization (albeit at 0 K), is more than sufficient to account for the experimentally observed unit cell expansion of the STO capacitors on LAO. Overall, our first-principles calculations suggest that change in formal layer charges at the interfaces of metal-insulator-metal heterostructures can strongly polarize the dielectric layer and provide a plausible explanation for the presence of polarity in our capacitors.

Specifically, we propose that the band offsets associated with the interfacial dipoles correspond to the spatially varying built-in fields that may be responsible for the observed PFM contrast in our films. The calculated band offsets of $\approx\pm0.7$ eV for the two possible orientations of the stoichiometric structure represent an upper limit for the built-in field variation across the sample. In practice, we would expect the interfaces to be of mixed termination on the macroscopic scale, in order to reduce the overall electrostatic energy, as is indeed found experimentally in $LaNiO_3/SrTiO_3$ heterostructures [53]. The domain-like features observed in our films may therefore correspond to regions with a preferential (rather than complete) termination of one type or the other, as schematically depicted in Figure 5 (g). Such regions will correspondingly have a reduced charge discontinuity and a reduced built-in field compared to that expected for uniformly terminated interfaces, giving a variety of possible intermediate states between +0.7 V and $-0.7$ V. Indeed, experimentally, the "domain" structure is only observable within a range of applied bias $\Delta V \approx 0.2$ V. The picture of spatially varying preferential interface terminations is also consistent with the observation of stripe-like contrast in Figure 3 (c).

Other potential sources of static, spatially varying internal fields that are able to polarize our STO layers include structural or chemical defect distributions. For example, flexoelectric polarization generated by the complex strain distribution around misfit dislocations can induce sizeable internal fields [54,55]. Given the low dislocation density in our 7 nm-thick STO films, confirmed by STEM, this mechanism is unlikely to be dominant in this film. However, for the 30 nm-thick STO film grown on LAO, which exhibits a dense array of misfit dislocations and significant strain relaxation, the flexoelectric effect may dominate. Trapped electronic charges or ionic defects, including oxygen vacancies, which can accumulate at defect sites such as misfit dislocation cores may also contribute to these internal fields [56].

Finally, we have deliberately ignored from our simple model the effect of non-equilibrium field-induced processes such as charge injection [57] or ionic electromigration [28,58] which can give rise to spurious electromechanical effects. While such dynamic processes are likely active in our samples at room temperature and play an important role in the electrical properties [43], the low DC bias required to achieve full "switching" in PFM (see Figure 3 (a)) suggests that these non-equilibrium effects are less relevant to the domain-like signals. Therefore, we propose that the observed electromechanical switching arises primarily from a separate quasi-static mechanism, while non-equilibrium processes will be more important at large fields and higher temperatures, although they are still expected to be active, even under small DC bias conditions, albeit at much slower rates.

## ***Conclusions***

In this work, we have revisited the properties of compressively-strained $SrTiO_3$, a system in which a stable out-of-plane ferroelectric phase at elevated temperatures has been theoretically predicted. We have presented a systematic investigation of the structural, electrical and electromechanical behavior of capacitors based on $SrTiO_3$ films under large compressive strains of up to −3%, incorporating $RENiO_3$ as electrode layers. Although our $SrTiO_3$ films show enhanced out-of-plane tetragonality and a room-temperature electromechanical response akin to that of polydomain ferroelectrics, variable temperature electrical measurements revealed no clear ferroelectric behavior down to 15 K. To account for the observed electromechanical behavior of the capacitors, we propose a mechanism based on spatially varying internal fields. Using first-principles calculations, we show that such internal fields may arise from charge discontinuities at interfaces between the formally charged layers of the $RENiO_3$ electrodes and the formally neutral layers of $SrTiO_3$. Developing methods for controlling the local interface termination would therefore enable the realization of highly stable nanoscale polar patterns with bespoke electrical and electromechanical responses. For example, a similar distribution of built-in fields in a ferroelectric thin film could give rise to antiferroelectric-like switching, providing an alternative route to engineering synthetic antiferroelectrics [39,59].

## ***Acknowledgments***

E. S., P. K., F. R., I. S., J. A. P, C. M. and P. Z. acknowledge funding received from the EU Horizon 2020 program under the Marie Skłodowska-Curie ITN action MANIC through grant agreement No. 861153. J. Í.-G. acknowledges the financial support from the Luxembourg National Research Fund through grant C24/MS/18979315/BELCOM. A.L. acknowledges EPSRC for part-funding of his studentship (Grant No. EP/T518001/1). This work made use of the facilities of the N8 Centre of Excellence in Computationally Intensive Research (N8 CIR) provided and funded by the N8 research partnership and EPSRC (Grant No. EP/T022167/1). The Centre is coordinated by the Universities of Durham, Manchester, and York. This work also made use of the Hamilton HPC services at Durham University. E. S., L. H. and P. Z. acknowledge the use of London Centre for Nanotechnology Atomic Force Microscope Facility and helpful discussions with Iurii Tikhonov and Igor Luk'yanchuk.

## ***Experimental methods***

### ***Epitaxial growth***

All heterostructures were grown in-situ via off-axis radiofrequency (RF) magnetron sputtering at 490°C. $SrTiO_3$ was grown in a 180 mTorr atmosphere with $Ar/O_2$ ratio of 28:20 at 60 W RF power, $NdNiO_3$ was grown in a 135 mTorr atmosphere with $Ar/O_2$ ratio of 30:10 at 50 W RF power, and $LaNiO_3$ was grown in a 110 mTorr atmosphere with $Ar/O_2$ ratio of 30:20 at 50 W RF power.

### ***Fabrication***

For the investigation of the out-of-plane electrical and electromechanical properties of the $SrTiO_3$-based heterostructures, we first deposited a thin Pt layer (~10 nm) on top using direct RF sputtering to improve the top contact quality, and then the samples were patterned using standard UV lithography into circular pads of various dimensions. Ar-ion milling was then used to etch the top Pt/$RENiO_3$ bilayer electrode and isolate the pads to create individual parallel-plate capacitors.

### ***Structural and electromechanical characterization***

X-ray measurements were performed with Cu K$\alpha_1$ radiation using a 9 kW Rigaku Smartlab diffractometer equipped with a rotating anode and a 2-bounce (220) Ge monochromator. Studies of the surface topography as well as the electromechanical experiments were performed with a Bruker Dimension Icon atomic force microscope operating in contact mode. Bruker's Si SCM-PIT-v2 (k=3 N/m) probes with Pt/Ir coated tips were used for the PFM measurements. For PFM in parallel-plate capacitor geometry, the sample was mounted on a metallic disc using silver paste and the bottom electrode was contacted by applying silver paste to the side walls of the sample. The top electrode was contacted using an indium wire held at the same potential as the scanning tip using a Bruker Signal Access Module (SAM), as depicted in the inset of Figure 3 (b). The electromechanical response was excited by applying an AC bias to the bottom electrode at a frequency near the resonance of the cantilever; the corresponding peak-to-peak amplitudes are detailed in the relevant figure captions. The DC bias was applied to the top electrode.

### ***Scanning transmission electron microscopy***

Cross-sectional lamella specimens were prepared using an FEI Helios 650 Dual Beam system. Scanning transmission electron microscopy (STEM) measurements in high-angle annular dark field (HAADF) mode were performed using a probe-corrected Thermo Fischer Scientific Titan 60-300 transmission electron microscope operated at 300 kV. The microscope is equipped with a high-brightness Schottky field emission gun (X-FEG), a Wien-filter monochromator, and a CETCOR corrector for the condenser system to provide sub-Å resolution in STEM mode. Geometric phase analysis of the HAADF-STEM images was carried out to determine the strain variations of the films, using the substrate as the lattice reference [60,61].

***Electrical characterization***

The low temperature electrical characterization (from 300 K down to 10 K) was performed in a Janis CCS-150 closed cycle cryostat. For the temperature-dependent dielectric characterization, an Agilent E4980A precision LCR meter was used to measure the complex impedance of the capacitor under study using an AC tickle bias of 10 mV in a frequency range from 100 Hz up to 2 MHz. The capacitance was subsequently calculated from the equivalent impedance value. The capacitance-voltage data correspond to 1 kHz measurements performed in $C_p$-D mode. The effective dielectric permittivity was calculated using $\varepsilon_r = Cd/\varepsilon_0 A$, where d is the thickness of the film, A is the capacitor area and $\varepsilon_0$ is the permittivity of free space. For polarization-voltage measurements, a Keysight 3350B waveform generator was used for the application of 1 kHz triangular voltage pulses, while the current response was measured using a FEMTO DLPCA-200 current amplifier and recorded using a Keysight MSOX 3054A oscilloscope.

***First-principles calculations***

We performed DFT calculations within the generalised gradient approximation (GGA), utilising the PBEsol exchange-correlation functional [62], and projector-augmented wave (PAW) generated pseudopotentials [63] within the PBE scheme [64], implemented in the Vienna Ab-initio Simulation Package (VASP) [65–67]. The Sr 4*s*, 4*p* and 5*s* (4*d* is also included, but initially left unoccupied); Nd 5*s*, 5*p*, 5*d*, and 6*s*; Ti 3*s*, 3*p*, 3*d* and 4*s*; Ni 3*p*, 3*d* and 4*s*; and O 2*s*, 2*p* electrons were explicitly included as valence, while all other electrons (including Nd 4*f*, assuming a 3+ ionisation state) were frozen in the ionic cores. We use a Hubbard-*U* correction of 2 eV on the Ni-3*d* sites. Both bulk-strained STO and interface calculations are based on a 20-atom perovskite cell with lattice vectors $a_{001} = a_p - b_p$, $b_{001} = a_p + b_p$, $c_{001} = 2c_p$, where $a_p$, $b_p$ and $c_p$ are primitive cubic cell vectors.

To investigate the effect of strain on STO, we undertook bulk-strained calculations on the 20-atom STO cell, where the in-plane lattice parameters are fixed to a chosen value, and the rest of cell relaxes. For these calculations, the plane-wave energy cut-off was set to 900 eV, while the K-point grid was fixed to a 5 x 5 x 4 Γ-centred mesh. To apply epitaxial strain, we fixed the in-plane components of the stress tensor using a stress-constraining geometry optimisation patch [68].

We use slightly different parameters for our NNO-STO interface calculations. For computational tractability during ionic relaxations, we set the plane-wave energy cut-off to 550 eV, while the K-point grid was fixed to 5 x 5 x 1 Γ-centred mesh for all systems. To simulate the film thicknesses used experimentally, while not exceeding computational limitations, we build interfaced systems using the same 20-atom perovskite cell as in the bulk-strained calculations. This is sufficient for simulating the metallic, ferromagnetic *Pbnm* structure of NNO, and *I*4*cm* (polar and tilted) or *I4/mcm* (non-polar and tilted) structure of STO. We remark that this does not allow for the simulation of the ground state insulating T-AFM phase of NNO [69,70], as it requires the doubling of the cell size in-plane. We relax all interfaced systems to a force convergence of 1 meV/Å. For all calculations, we fix the in-plane lattice parameters to those of LAO. To obtain these parameters, we relaxed [001]-oriented bulk LAO (including La 4*f* (unoccupied), 5*s*, 5*d*, 5*p* and 6*s*; Al 3*s* and 3*p* electrons) in its ground state rhombohedral $R\bar{3}c$ phase ($a^-a^-a^-$ tilt pattern), and found the result to be comparable to low-temperature neutron diffraction data [71], such that we obtained in-plane lattice parameters of $|a^{LAO}_{001}| = 5.35$ Å and $|b^{LAO}_{001}| = 5.34$ Å.

We relax three systems in total. The two non-stoichiometric interfaced systems consist of two of the same type of interface, either SrO-$NiO_2$ or NdO-$TiO_2$. This fixes net inversion symmetry if starting from a non-polar STO phase (though it is still broken locally at interfaces), preventing STO from accessing the polar phase. The system consisting of two SrO-$NiO_2$ interfaces has 8.5 NNO layers and 11.5 STO layers, while the system of two NdO-$TiO_2$ interfaces has 9.5 NNO layers and 10.5 STO layers (where each layer has the chemical formula $2ABO_3$). The stoichiometric system, made up of 8 NNO layers and 12 STO layers (where each layer is also $ABO_3$), consists of one of each interface termination.

## Supplementary Information

# Alternative origins of polarity in compressively strained $SrTiO_3$-$RENiO_3$ capacitors

Evgenios Stylianidis[1,2], Panagiotis Koutsogiannis[3,4], Alexander Lione[5], Felix Risch[6], Laura Hechler[1,2], Igor Stolichnov[6], Jorge Íñiguez-González[7,8], José A. Pardo[3,9], Nicholas C. Bristowe[5], César Magén[3,4], Pavlo Zubko[1,2]

[1] Department of Physics and Astronomy, University College London, Gower Street, London WC1E 6BT, United Kingdom
[2] London Centre for Nanotechnology, 17-19 Gordon Street, London WC1H 0AH, United Kingdom
[3] Instituto de Nanociencia y Materiales de Aragón (INMA), CSIC-Universidad de Zaragoza, 50009 Zaragoza, Spain
[4] Departamento de Física de la Materia Condensada, Universidad de Zaragoza, 50018 Zaragoza, Spain
[5] Condensed Matter Physics, Physics Department, Durham University, Durham, United Kingdom
[6]Nanoelectronic Devices Laboratory (NanoLab), Ecole Polytechnique Fédérale de Lausanne (EPFL), 1015, Lausanne, Switzerland
[7] Smart Materials Unit, Luxembourg Institute of Science and Technology (LIST), Avenue des Hauts-Fourneaux 5, L4362, Esch-sur-Alzette, Luxembourg
[8] Department of Physics and Materials Science, University of Luxembourg, 41 Rue du Brill, L-4422 Belvaux, Luxembourg
[9] Departamento de Ciencia y Tecnologia de Materiales y Fluidos, Universidad de Zaragoza, 50018 Zaragoza, Spain

## **Supplementary Section A:** Phase diagram of compressively-strained $SrTiO_3$

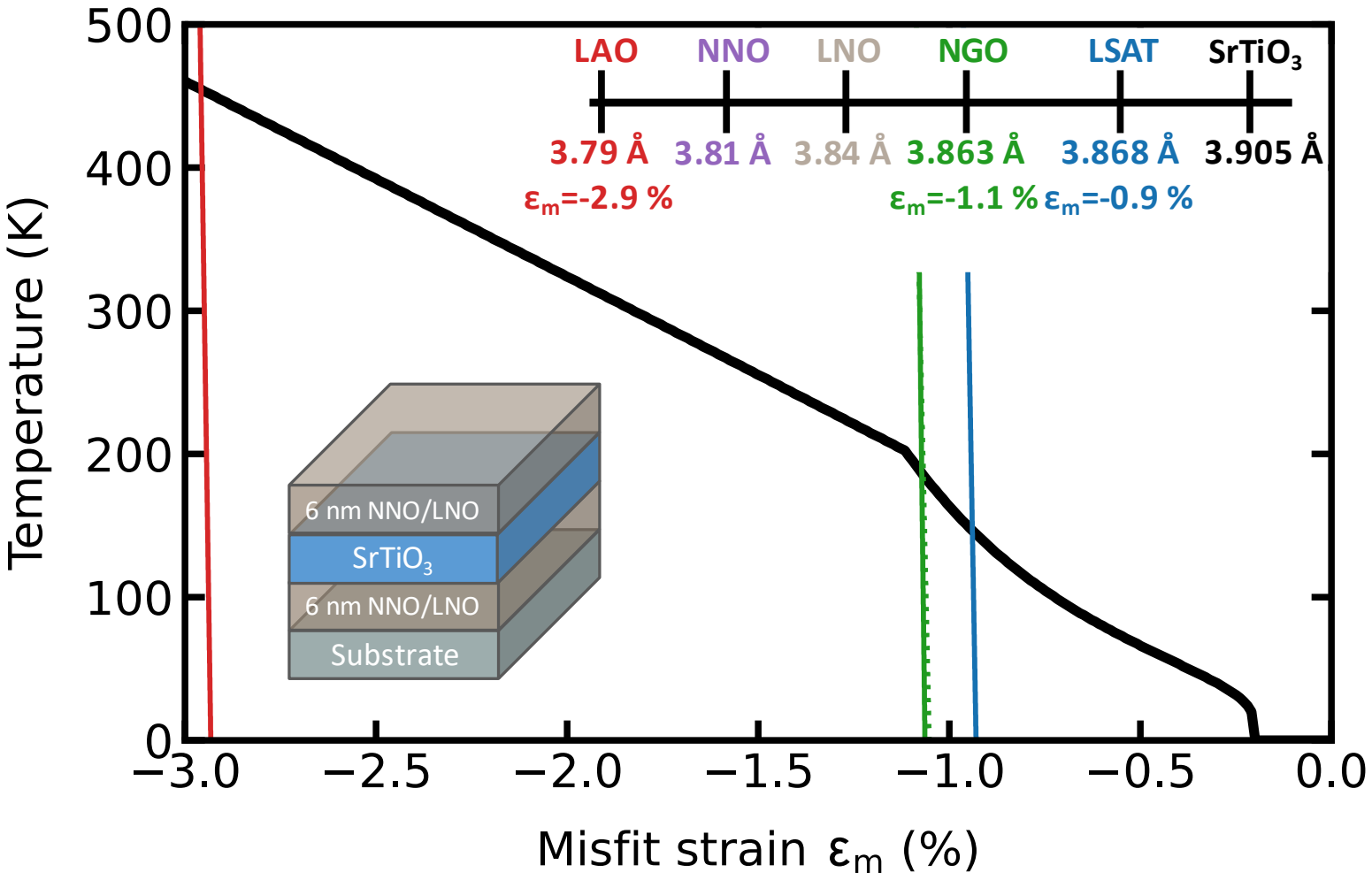

**Supplementary Figure A1:** Phase diagram of $SrTiO_3$. The black continuous line depicts the ferroelectric transition temperature as a function of misfit strain for $SrTiO_3$ as calculated by Pertsev et al.[1]. The continuous red, green and blue lines correspond to the misfit strain induced by $LaAlO_3$, $NdGaO_3$ and LSAT, respectively, taking into account the thermal expansion coefficients (TEC) of each material. For $NdGaO_3$, the dashed and continuous lines account for the slightly different TECs of the substrate in the two in-plane directions. Top right inset: the materials used in the present study, their room-temperature lattice parameters, and the room-

temperature misfit strain with $SrTiO_3$ (not to scale). Bottom left inset: schematic of the $RENiO_3/SrTiO_3/RENiO_3$ heterostructures explored in this work.

## **Supplementary Section B:** Electrical properties of $RENiO_3$ thin films

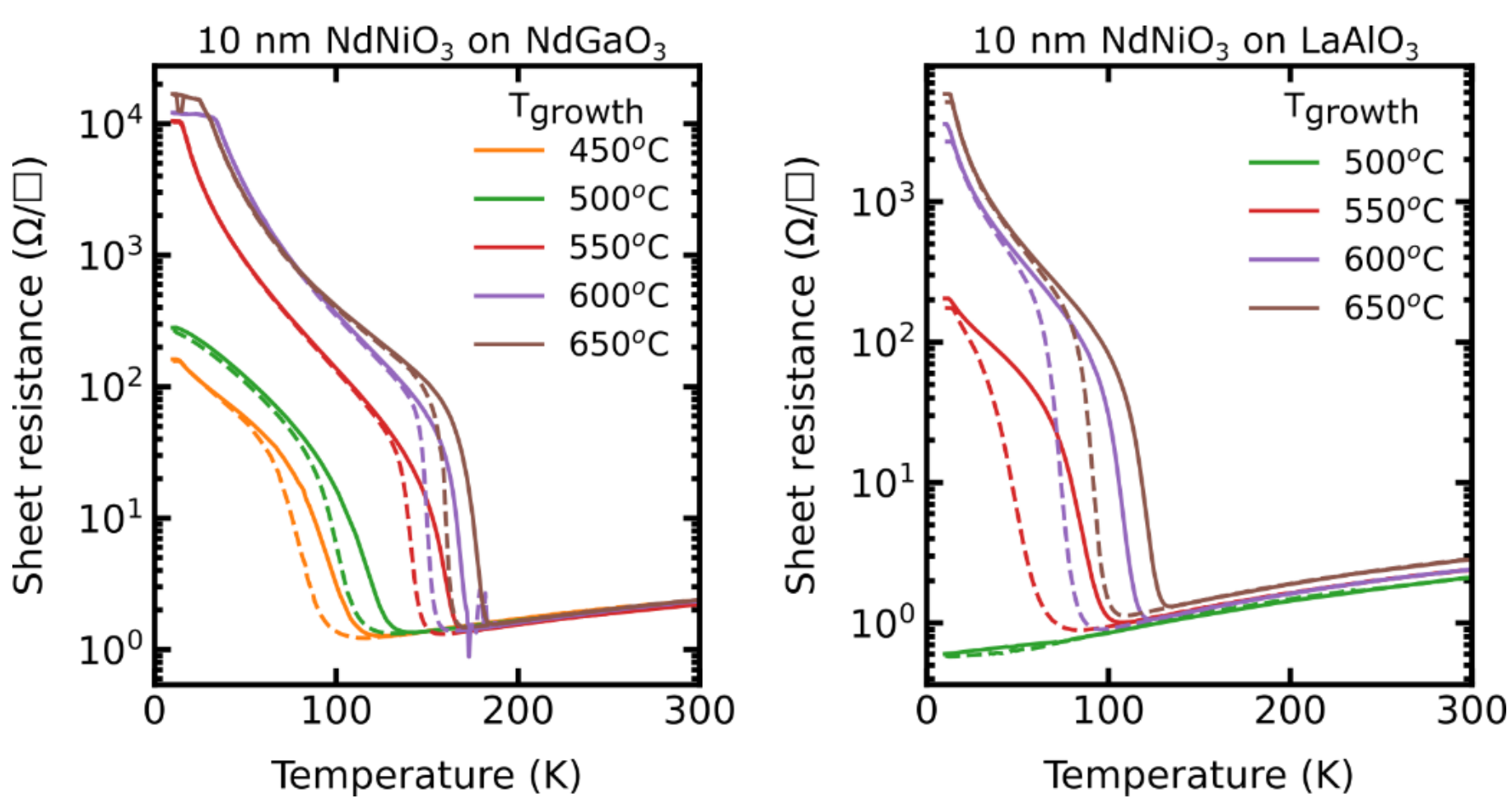

**Supplementary Figure B1:** Sheet resistance as a function of temperature for 10 nm $NdNiO_3$ films grown on $110_o$-oriented $NdGaO_3$ and $001_{pc}$-oriented $LaAlO_3$ substrates grown at different temperatures. Continuous (dashed) lines represent heating (cooling) legs of the temperature cycles.

## **Supplementary Section C:** The effect of the electrode MIT on impedance spectroscopy measurements

The thermally-driven metal-insulator transition (MIT) in $RENiO_3$ compounds (except for $LaNiO_3$, which remains metallic at all temperatures) provides a mechanism to study the effect of charge screening on the spontaneous polarization of ferroelectric capacitors employing $RENiO_3$ electrodes. Although this aspect is beyond the scope of the present study, Supplementary Figures C1 and C2 show the effect of the electrode MIT on impedance spectroscopy measurements.

Supplementary Figure C1 compares the frequency and temperature dependence of the real (C') and imaginary (C'') parts of the complex capacitance, obtained from the measured complex impedance using $C^* = 1/i\omega Z^*$, for two types of capacitors grown on LSAT: one with $NdNiO_3$ as top and bottom capacitor plates, and one with $LaNiO_3$. In both cases, the C' value in the plateau region decreases with temperature. The primary difference in the behavior of the two samples is the significant shift of the drop in C' and the maximum in C'' towards lower frequencies with decreasing temperature for the $NdNiO_3$ capacitors, indicating an increase in the series resistance of the system. To extract quantitative information, we fitted the frequency sweeps to a simple equivalent circuit model (inset of Supplementary Figure C1 (a)) consisting of a resistor $R_s$, representing the series resistance contribution from the capacitor electrodes and connections, in series with a capacitor $C_{STO}$, the capacitance related to the dielectric response of $SrTiO_3$. Although more complex models could lead to better fitting of our data, this simple model captures well the main features of the frequency sweeps (the plateau in C', the drop in C' and the position of the C'' maximum).

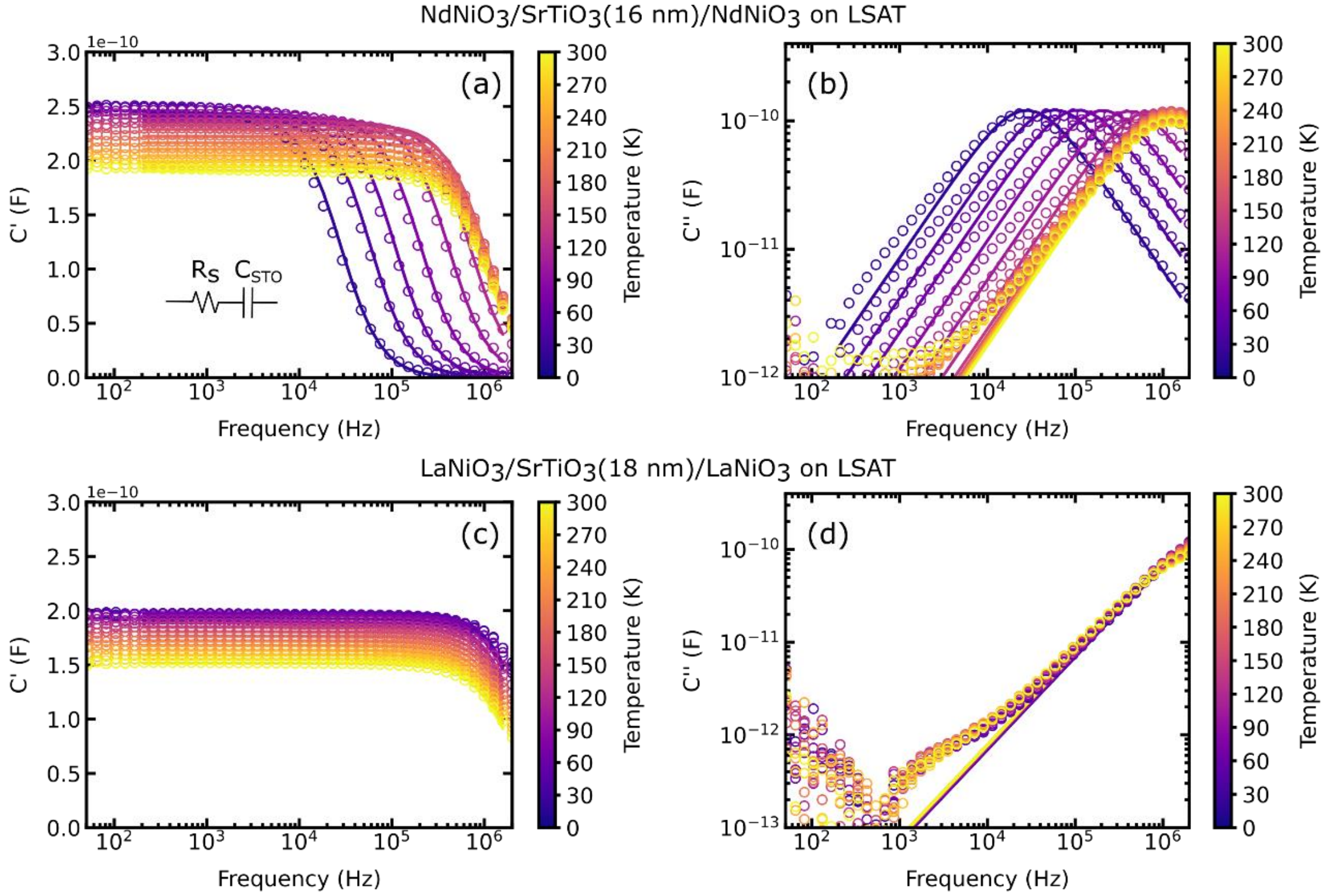

**Supplementary Figure C1:** Frequency and temperature dependence of the measured real and imaginary parts of the capacitance for Pt-coated $NdNiO_3/SrTiO_3$(16 nm)/$NdNiO_3$ (a and b) and $LaNiO_3/SrTiO_3$(18 nm)/$LaNiO_3$ (c and d) heterostructures grown on LSAT. The open circles represent raw data, while the continuous lines are fits to the equivalent circuit depicted in the inset of (a). The measurements are from circular capacitors with 25 μm radius with a 10 mV excitation bias.

The fitted parameters are plotted in Supplementary Figure C2. Error bars obtained from the fit for each data point are also included although occasionally they are difficult to discern due to the small error. For the $LaNiO_3$-based system, $R_s$ decreases almost monotonically with decreasing temperature showing a metallic behavior. By contrast, for the system with $NdNiO_3$, $R_s$ clearly transitions from a metallic to an insulating behavior at about 130 K, resembling the expected MIT of $NdNiO_3$. The series resistance $R_s$ includes contributions from the resistance of the external wires and contacts (which is not expected to have a strong temperature dependence), the sheet resistance of the bottom $NdNiO_3$ and the resistance of the top $NdNiO_3$-Pt bilayer. The conductivity of the $NdNiO_3$-Pt bilayer will be dominated by the conductivity of Pt; hence we expect no significant change in resistance of the top contact across the $NdNiO_3$ MIT. Therefore, the main temperature dependence of the extracted $R_s$ is expected to be due to the temperature dependence of the sheet resistance of the bottom $NdNiO_3$ layer.

The fitted capacitance values $C_{STO}$ exhibit a maximum around 30 K for both samples. An additional weak anomaly in $C_{STO}$ is observed for the $NdNiO_3$-based sample near the MIT temperature. However, this anomaly is less obvious in other samples with an MIT (e.g. those grown on $NdGaO_3$) and requires further investigation. The main role of the MIT of the bottom electrode is therefore to increase the series resistance and shift the RC time constant related features to lower frequencies, reducing the accessible frequency window for dielectric measurements.

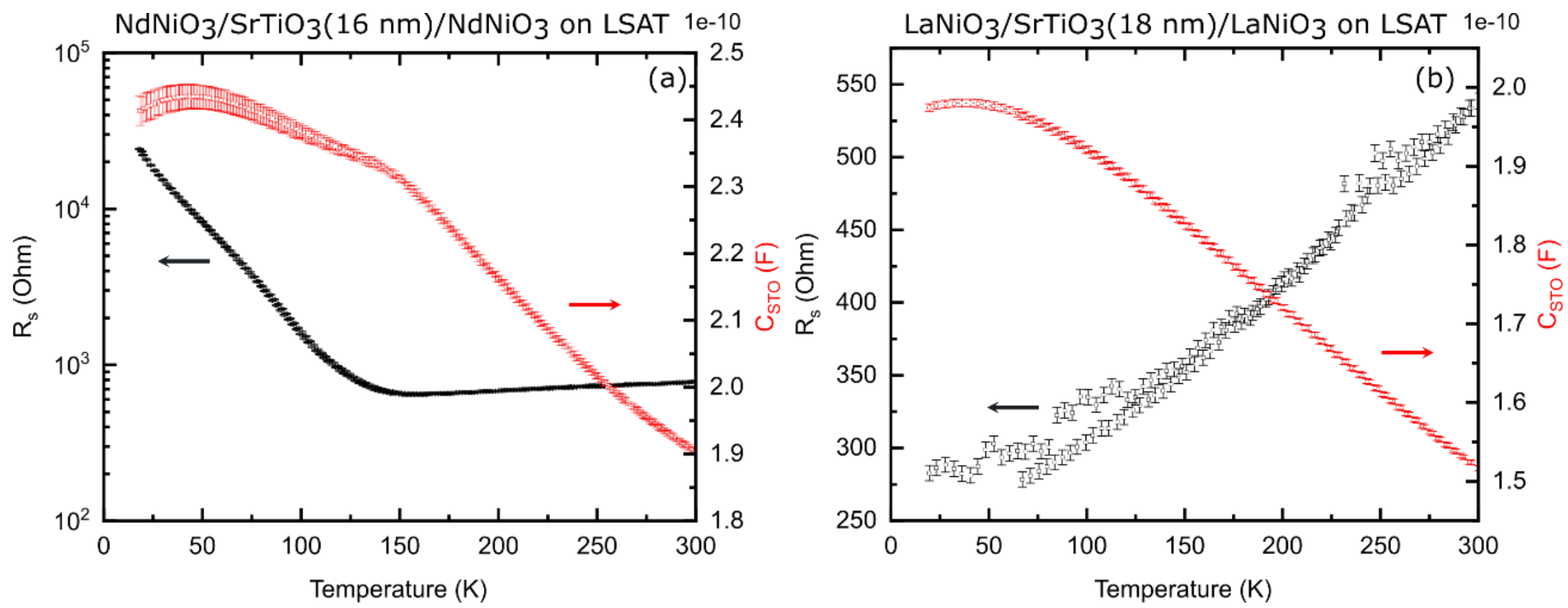

**Supplementary Figure C2:** Fitting parameters $R_s$ and $C_{STO}$ obtained from the fits shown in Supplementary Figure C1 for (a) $NdNiO_3$/$SrTiO_3$(16 nm)/$NdNiO_3$ and (b) $LaNiO_3$/$SrTiO_3$(18 nm)/$LaNiO_3$ heterostructures grown on LSAT.

## <u>Supplementary Section D:</u> Additional structural characterization

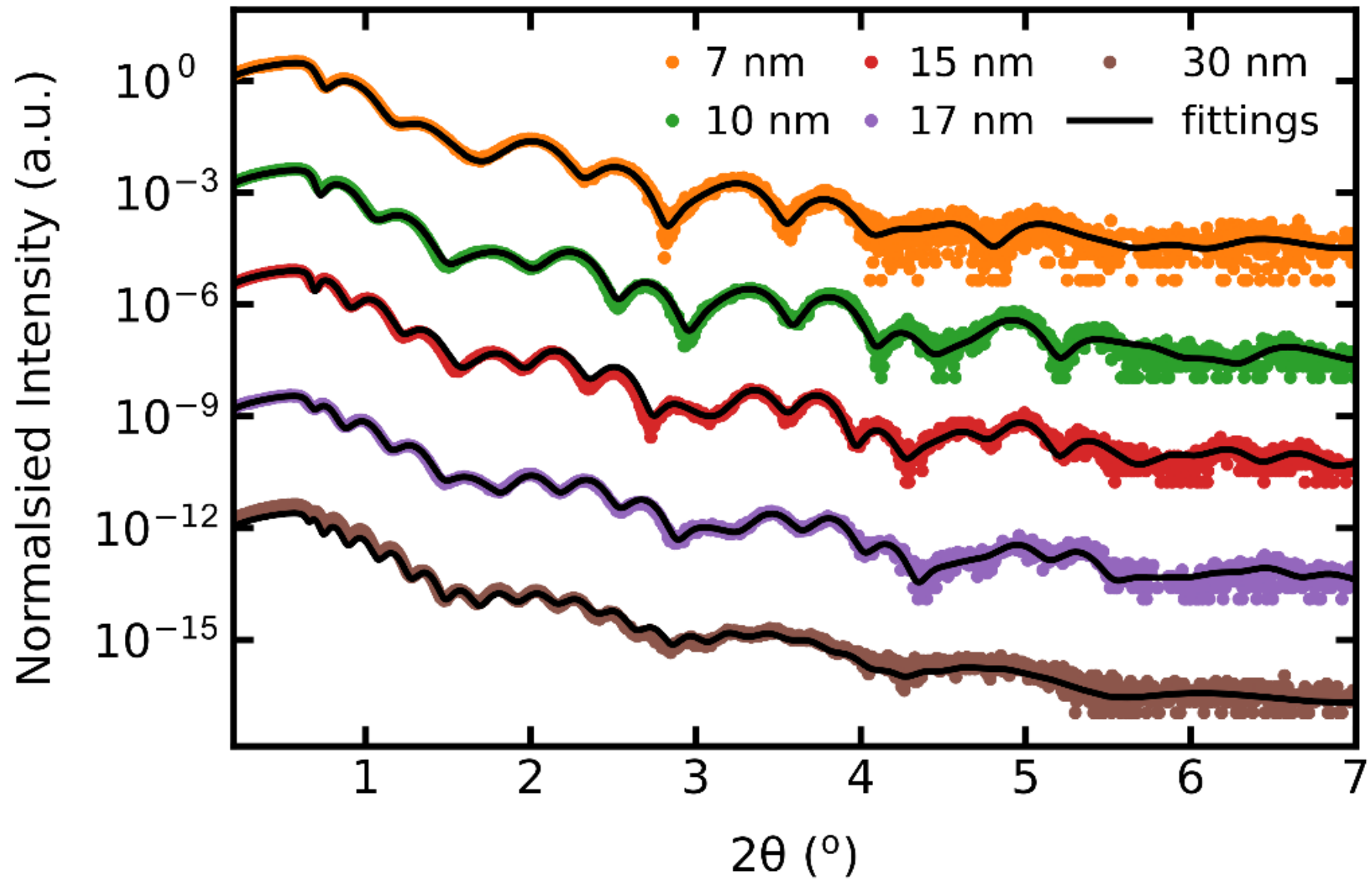

**Supplementary Figure D1:** X-ray reflectivity θ-2θ measurements for $NdNiO_3$/$SrTiO_3$/$NdNiO_3$ heterostructures with various thicknesses of $SrTiO_3$ grown on $LaAlO_3$. Fits were performed using the *GenX* software[2].

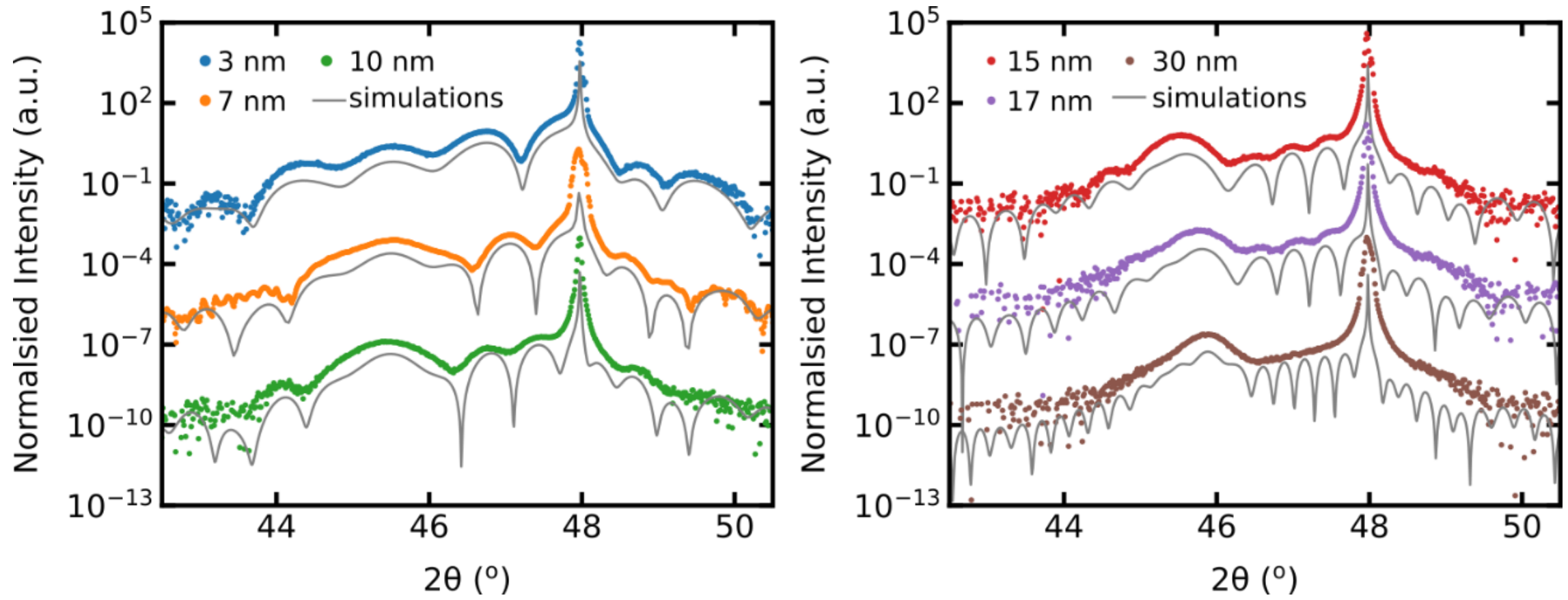

**Supplementary Figure D2:** X-ray diffraction θ-2θ measurements around the substrate $002_{pc}$ peak for $NdNiO_3/SrTiO_3/NdNiO_3$ heterostructures grown on $LaAlO_3$ for various $SrTiO_3$ thicknesses. The simulations were performed using the method described in [3].

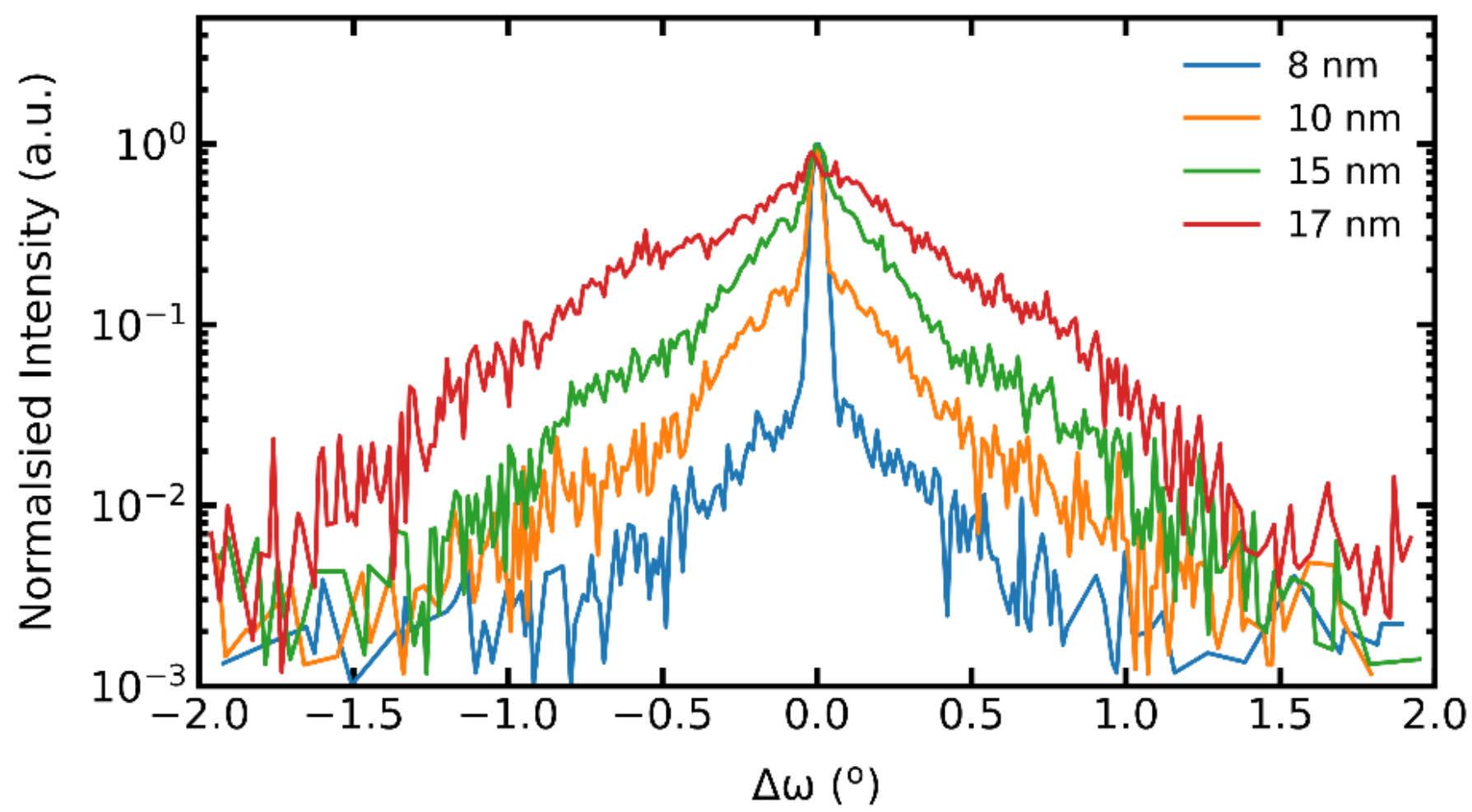

**Supplementary Figure D3:** Rocking curves around the $SrTiO_3$ $002_{pc}$ peak for $NdNiO_3/SrTiO_3/NdNiO_3$ heterostructures grown on $LaAlO_3$ for various $SrTiO_3$ thicknesses. The 8 nm curve is for a $SrTiO_3$ film directly grown on $LaAlO_3$.

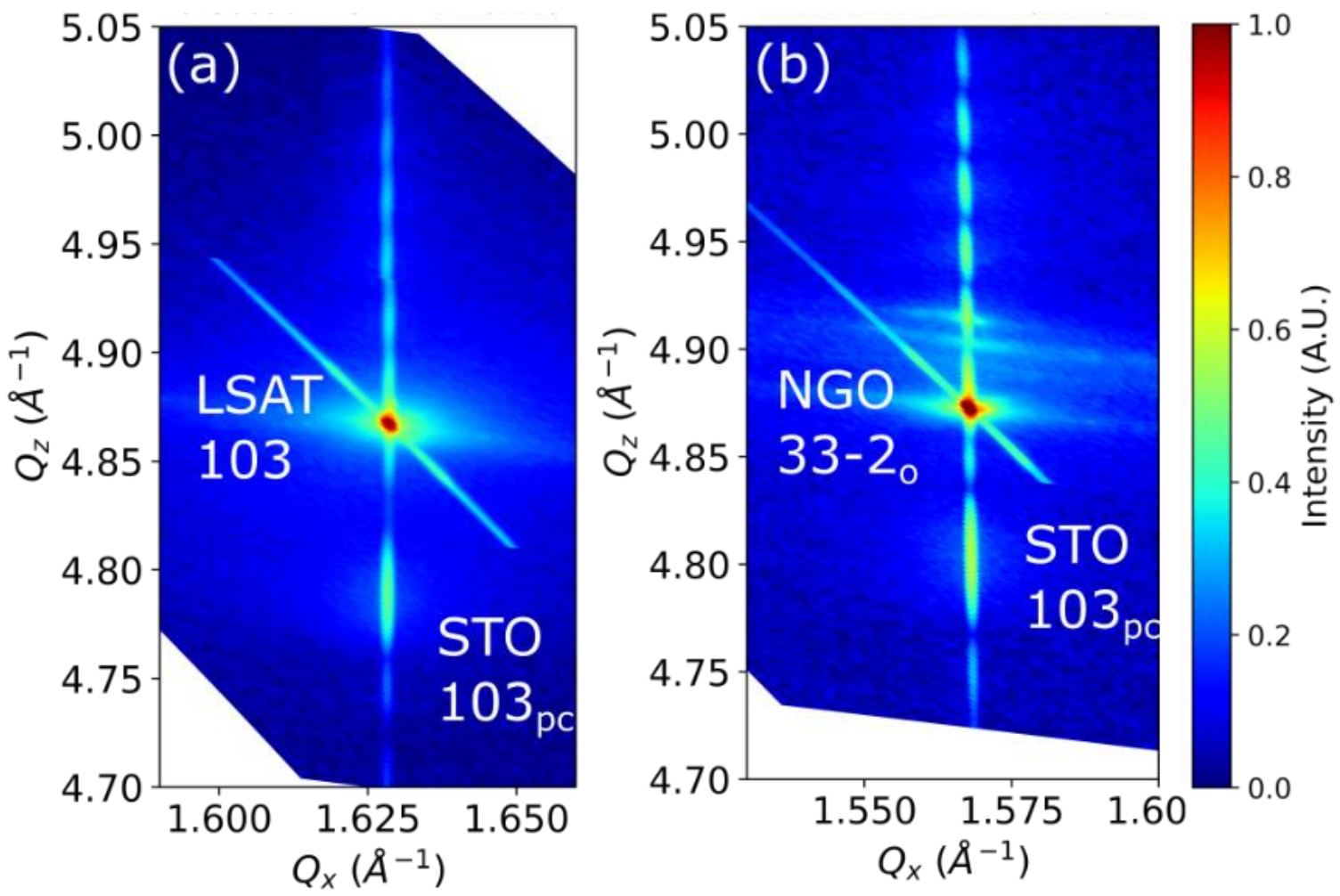

**Supplementary Figure D4:** Reciprocal space maps around the off-specular $103_{pc}$ (or equivalent) reflection for $LaNiO_3/SrTiO_3/LaNiO_3$ heterostructures with 30 nm $SrTiO_3$ grown on (a) LSAT and (b) $NdGaO_3$.

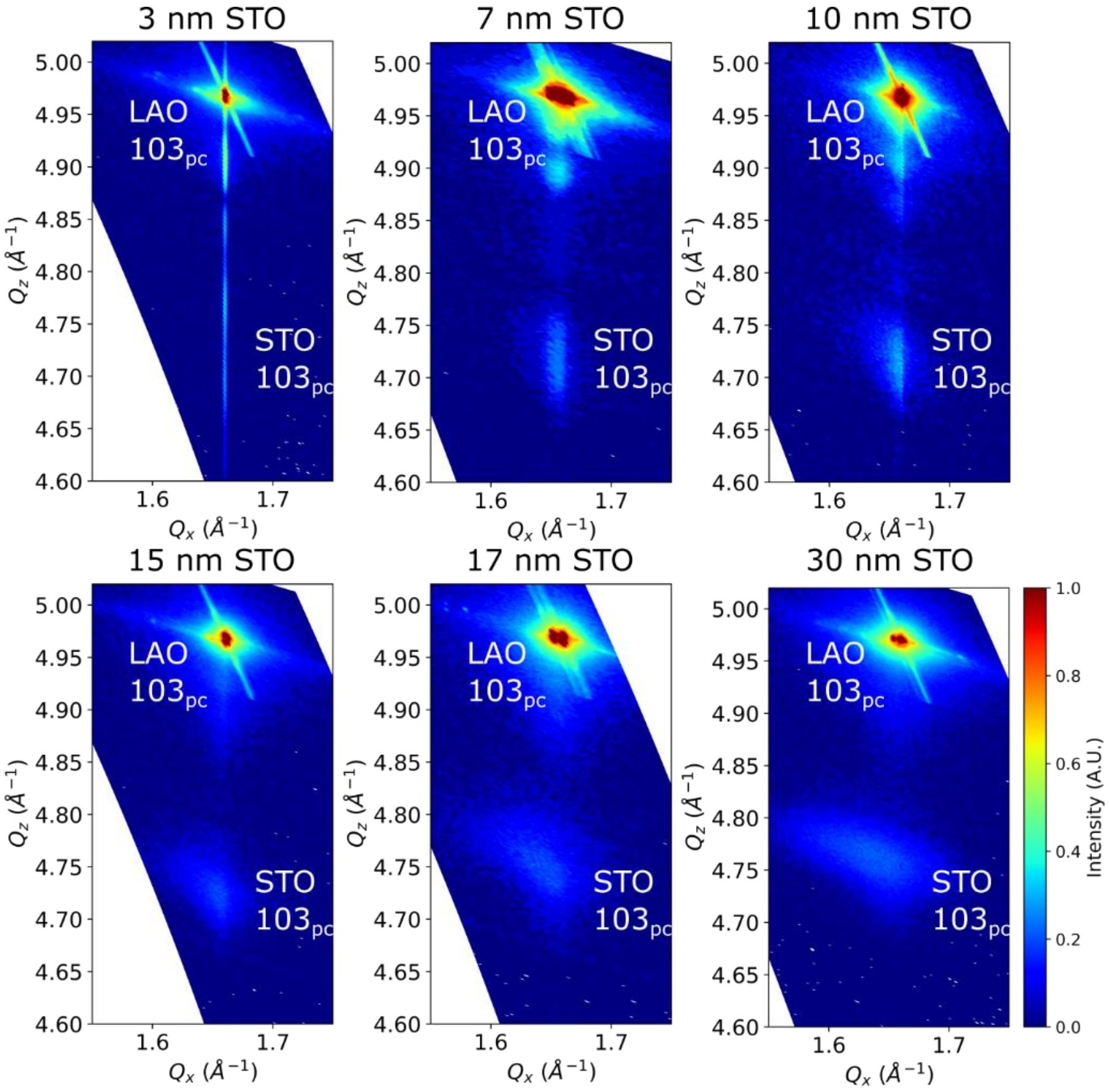

**Supplementary Figure D5:** Reciprocal space maps around the off-specular $103_{pc}$ (or equivalent) reflection for $NdNiO_3/SrTiO_3/NdNiO_3$ heterostructures with various $SrTiO_3$ thicknesses grown on $LaAlO_3$.

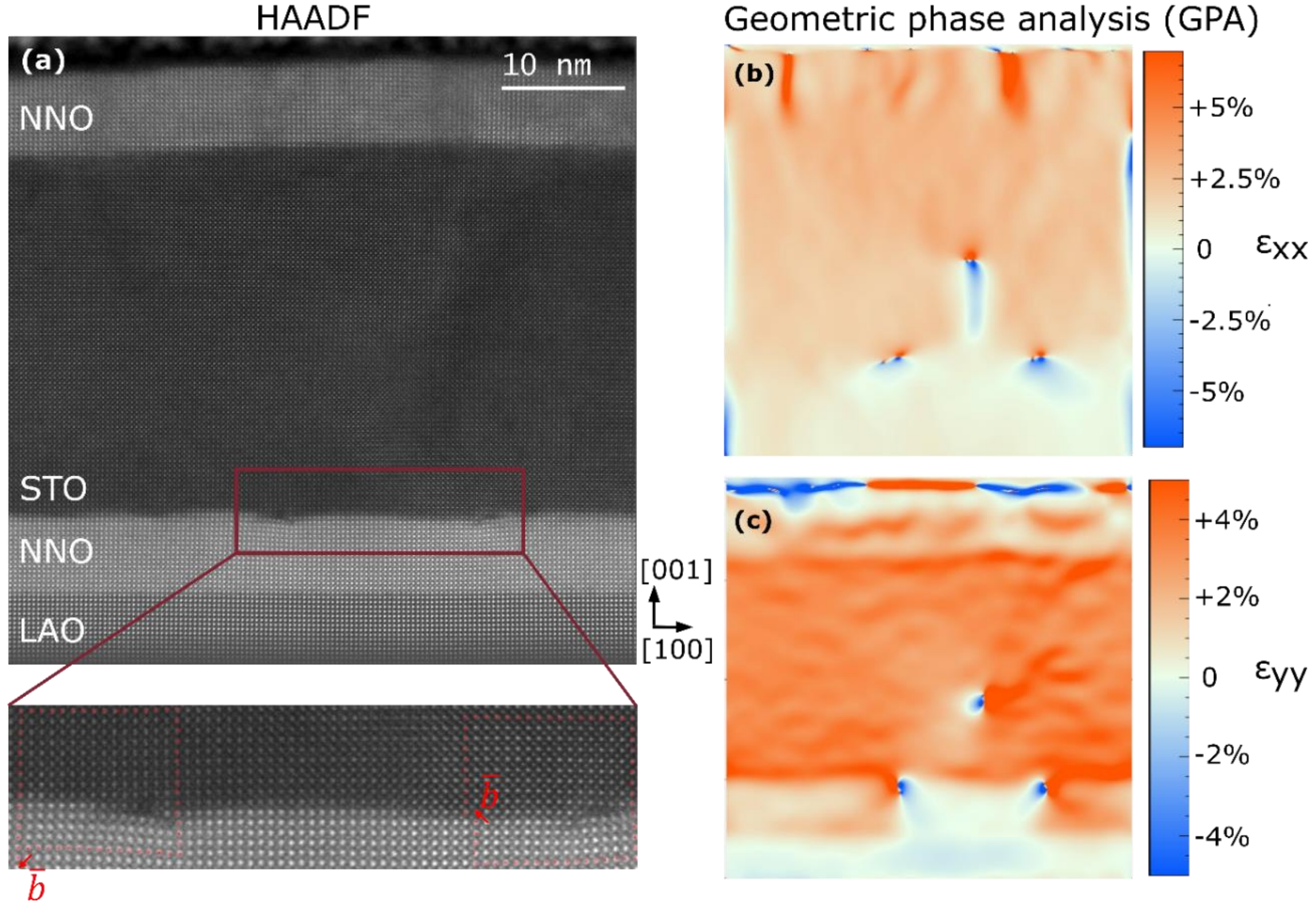

**Supplementary Figure D6:** STEM analysis of the $NdNiO_3/SrTiO_3$(30 nm)/$NdNiO_3$ heterostructure grown on $LaAlO_3$. (a) Atomic resolution HAADF-STEM cross-sectional image of the heterostructure. The dark red rectangle shows a magnified image of the bottom $NdNiO_3/SrTiO_3$ interface where two distinct dislocations are residing. The average Burgers vectors $\bar{b}$ for each dislocation are also shown (b) and (c) show in-plane $\varepsilon_{xx}$ and out-of-plane $\varepsilon_{yy}$ deformation maps, respectively, obtained after geometric phase analysis of the HAADF-STEM image in (a).

## **<u>Supplementary Section E:</u>** Additional electrical characterization

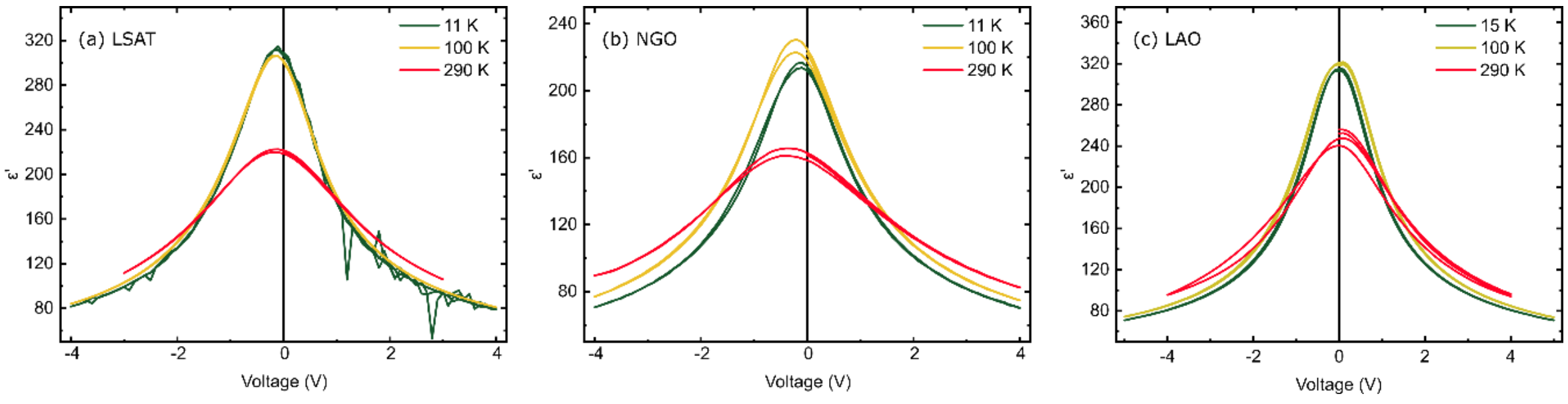

**Supplementary Figure E1:** Permittivity-voltage measurements for Pt-coated $RENiO_3/SrTiO_3/RENiO_3$ heterostructures with 30 nm $SrTiO_3$ grown on (a) LSAT, (b) $NdGaO_3$ and (c) $LaAlO_3$.

## **Supplementary Section F:** Additional PFM measurements

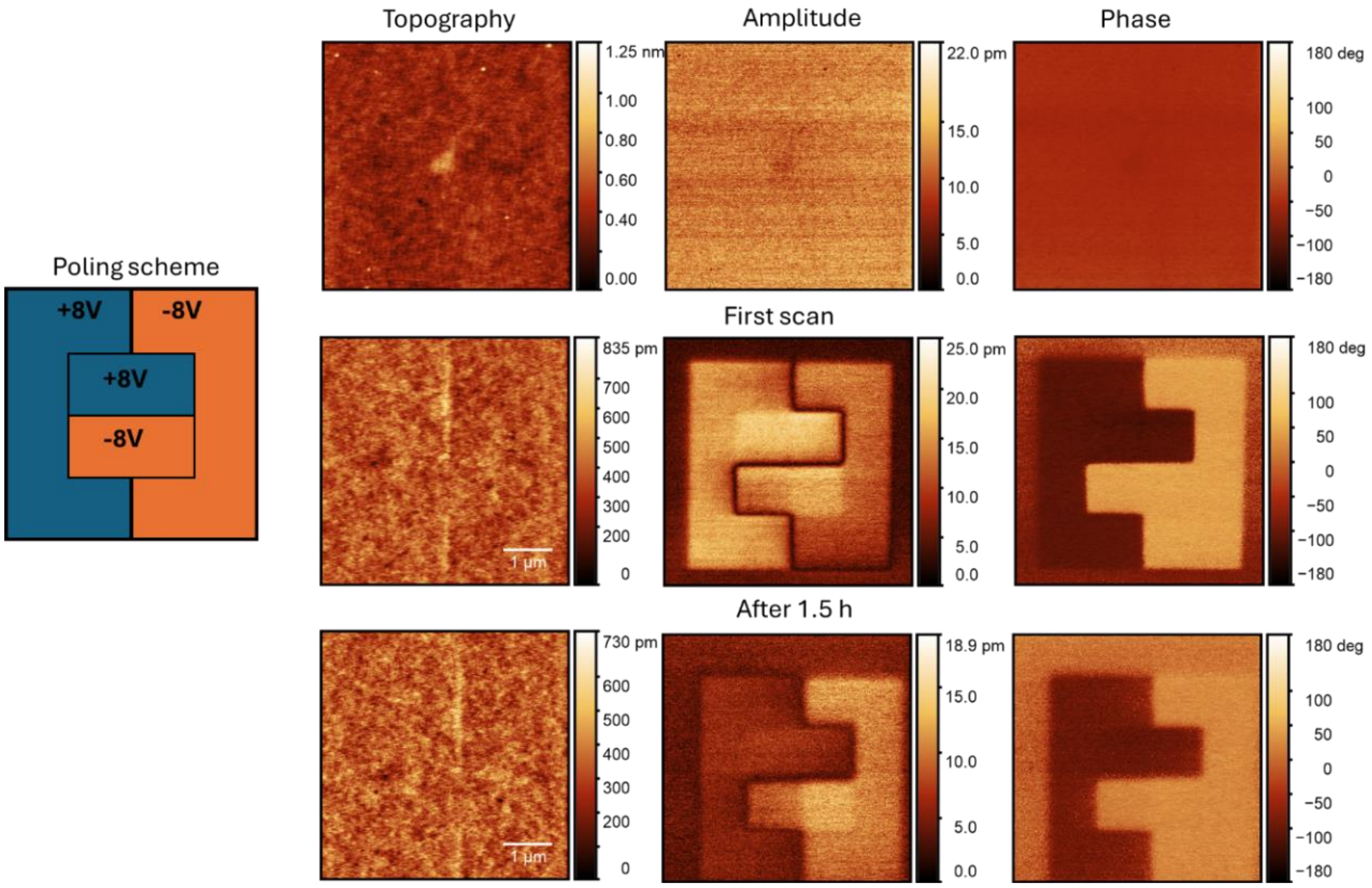

**Supplementary Figure F1:** Vertical PFM switching experiments on the bare surface of a 7 nm thick $SrTiO_3$ film grown on $LaAlO_3$ buffered with a 6 nm thick conductive $NdNiO_3$ layer. The first row shows topography, PFM amplitude and PFM phase images of the as-grown sample. Then, the DC poling pattern shown on the left was applied to the samples. The second row shows data immediately after poling, while the data on the third row were recorded 1.5 hours after poling.

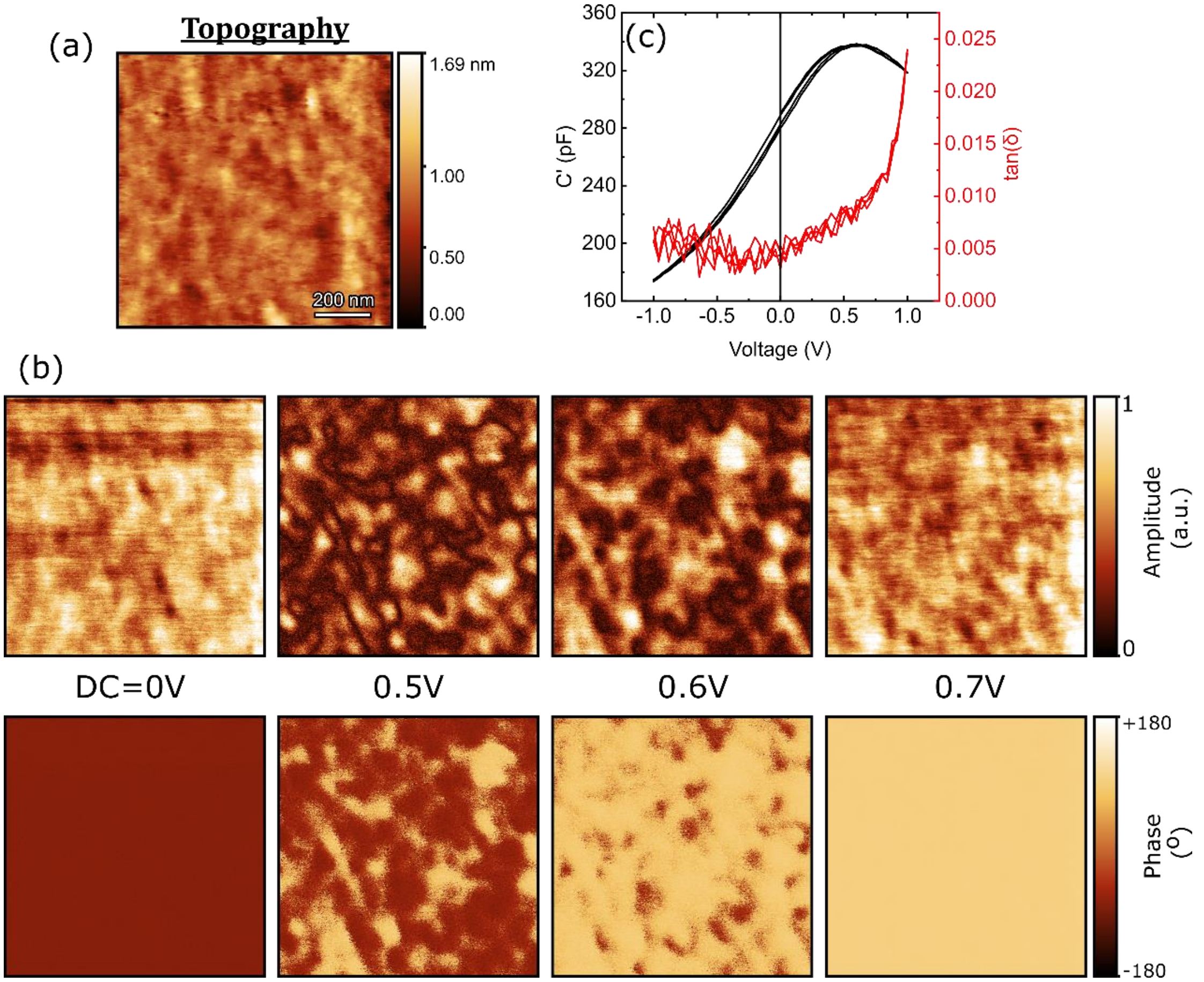

**Supplementary Figure F2:** (a) Topography and (b) DC-bias dependent vertical PFM measurements through the top electrode on a Pt-coated pad with 50 μm radius of α $LaNiO_3/SrTiO_3$(7 nm)/$LaNiO_3$ heterostructure grown on $LaAlO_3$. The measurements were performed with a 1 V and 245 kHz AC bias. (c) Representative capacitance-voltage and loss-tangent-voltage curves for the same sample but a different pad.

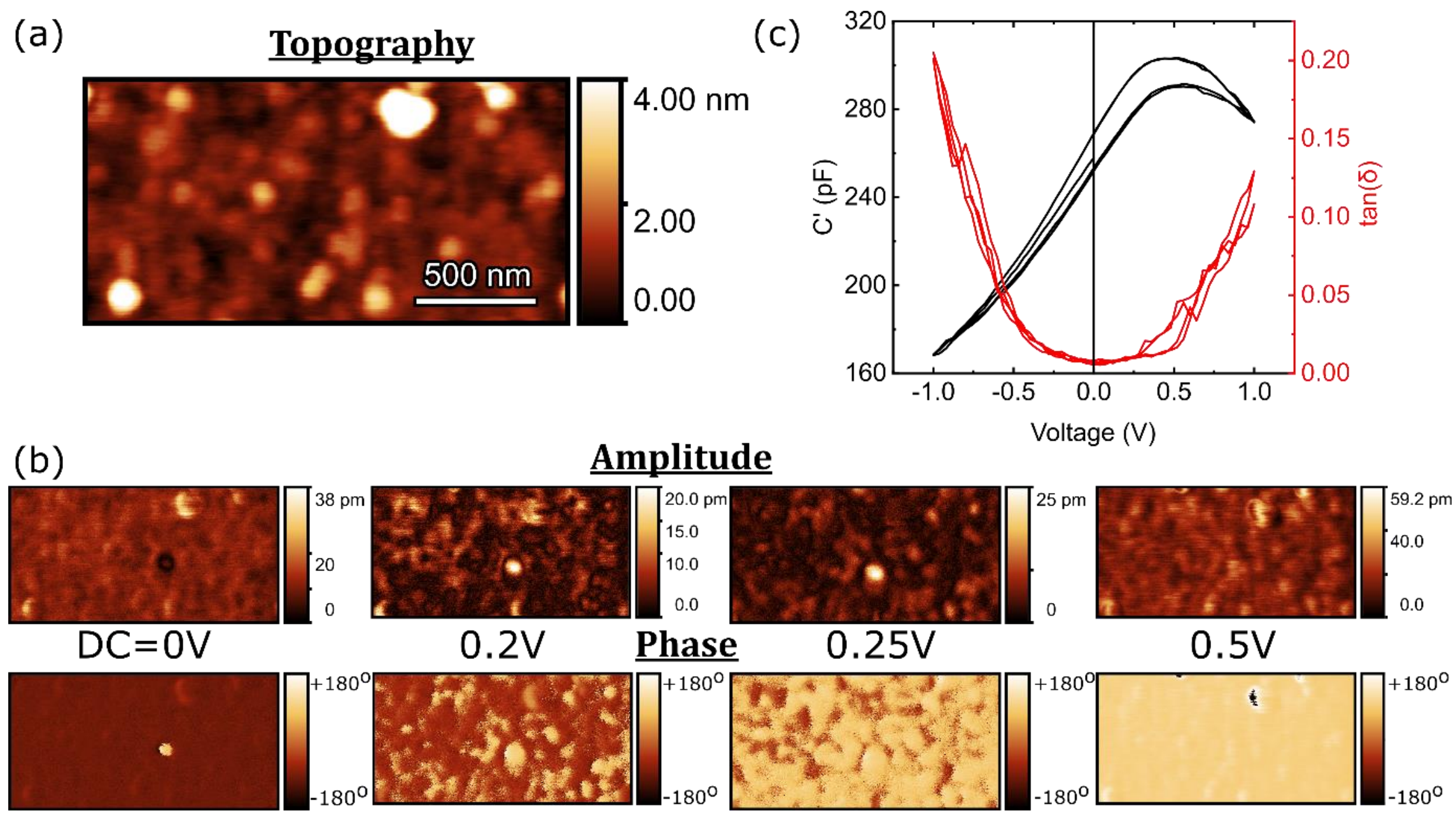

**Supplementary Figure F3:** (a) Topography and (b) DC-bias dependent vertical PFM measurements through the top electrode on a Pt-coated pad with 50 μm radius of the $LaNiO_3/SrTiO_3$(7 nm)/$LaNiO_3$ heterostructure grown on $NdGaO_3$. The measurements were performed with a 1 V and 260 kHz AC bias. (c) Representative capacitance-voltage and loss-tangent-voltage curves for the same sample but a different pad.

## **Supplementary Section G:** Discussion on excess lattice expansion of $SrTiO_3$ films

From the experimentally determined in-plane and out-of-plane lattice parameters of our $SrTiO_3$ films (Figure 1 (d) in the main text), it is evident that our strained films exhibit excess out-of-plane lattice expansion compared to the theoretical estimates. A plausible reason for the observed unit cell expansion in our films is the development of spontaneous strains associated with the development of an order parameter, either a spontaneous out-of-plane polarization $P_z$, or the AFD order parameter with out-of-plane rotation axis $q_z$. According to the phase diagram of Pertsev et al. [1], at room temperature, $SrTiO_3$ films strained on $LaAlO_3$ are expected to be in the AFD phase, while films on $NdGaO_3$ should be paraelastic. Using a thermodynamic method, the tetragonality of $SrTiO_3$ can be expressed as:

$$c_{film}/a_{film}=(1-\varepsilon_m)(1+[g_{11}P_z^2+\lambda_{11}q_z^2-2c_{12}\varepsilon_m]/c_{11}) \quad (1)$$

where $g_{11}=1.25\text{x}10^{10}$ Jm/C$^2$, $\lambda_{11}=1.3\text{x}10^{30}$ J/m$^5$, $c_{12}=1.07\text{x}10^{11}$ J/m$^3$ and $c_{11}=3.36\text{x}10^{11}$ J/m$^3$ [1]. We use the expression above to estimate the polarization or AFD rotation angle required to match the experimental tetragonality of our film. For the 7 nm $SrTiO_3$ film on $LaAlO_3$, $c_{film}/a_{film}=1.055$, $q_z=0$ yields $P_z=66$ μC/cm$^2$, while for $P_z=0$ we estimate $q_z=65$ pm, which corresponds to $\phi_z=19^o$ [4]. A combination of polarization and oxygen octahedron rotations between these values could therefore, in principle, account for the observed tetragonality.

For the 30 nm $SrTiO_3$ film on $NdGaO_3$ we estimated $c_{film}/a_{film}=1.02$. Assuming it is paraelastic ($q_z=0$) at room temperature, equation (1) would imply $P_z=28$ μC/cm$^2$. However, neither theoretical predictions nor experimental studies have demonstrated ferroelectricity in this system at room temperature. Therefore, our analysis suggests an alternative origin of the enhanced out-of-plane expansion of at least some of our films.

Chemical expansion due to off-stoichiometry is a common effect in perovskite oxides. When intentional cationic off-stoichiometry was introduced in homoepitaxial $SrTiO_3$ films, an out-of-plane expansion of the lattice parameters of the films has been reported [5,6]. To investigate this further we grew a homoepitaxial $SrTiO_3$ film using the same conditions as our heteroepitaxial layers. X-ray characterization revealed an expanded unit cell, with a lattice parameter of 3.93 Å. This suggests that chemical expansion is a likely origin for the expanded unit cell in our films. We note that a direct comparison between the out-of-plane expansion of the homoepitaxial film and the strained films is not trivial, since the formation energies of different ionic vacancies are highly dependent on strain [7,8]. Although we intentionally grew all films under the same conditions, the optimal growth conditions will likely vary with the substrate and the imposed strain.

## **Supplementary Section H:** Electromechanical response with an external applied field

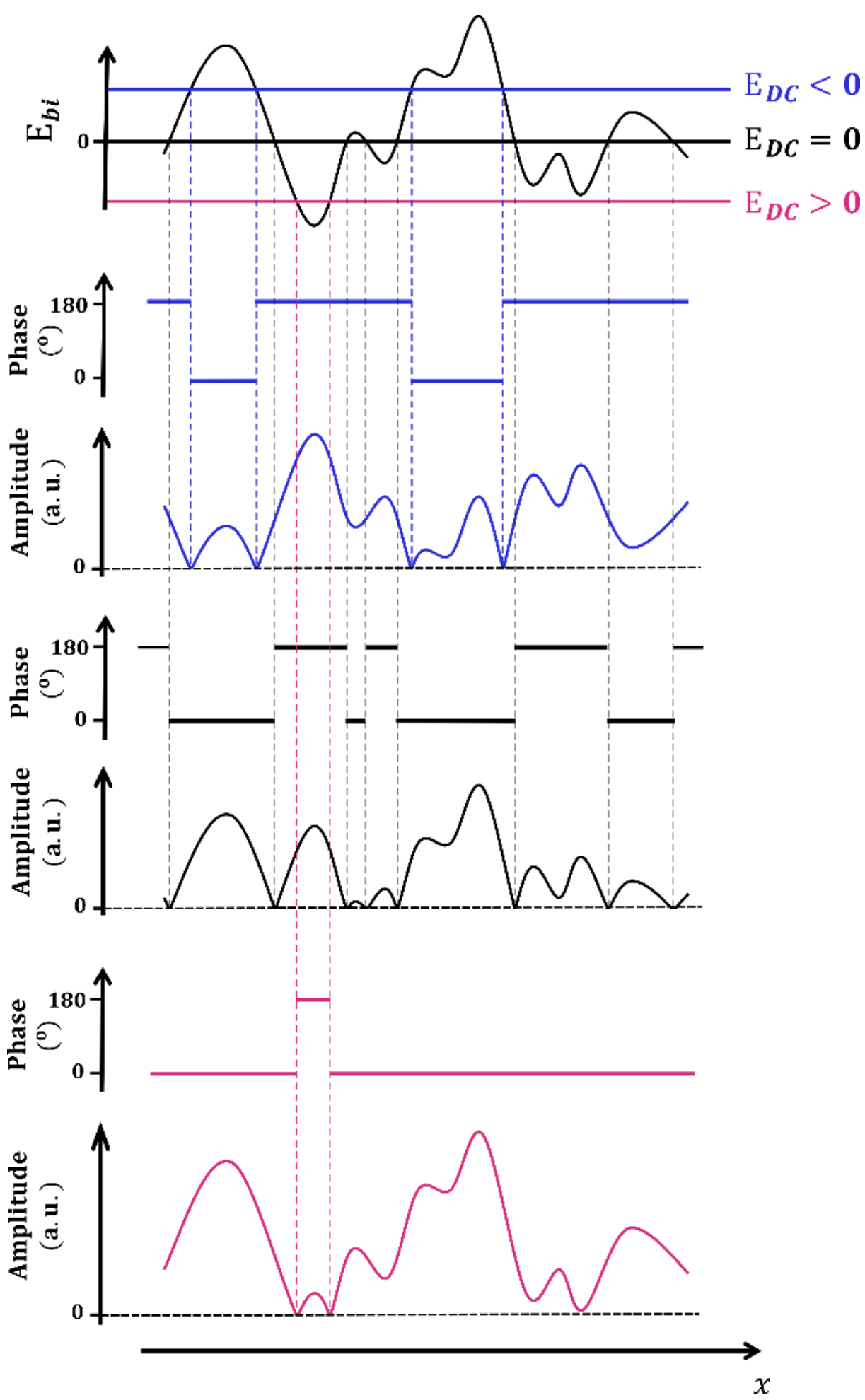

**Supplementary Figure H1:** Electromechanical response model. Schematic depiction of how a spatially varying built-in field can give rise to an electromechanical response reminiscent of a polydomain ferroelectric. The expected evolution of the PFM phase and amplitude signal upon application of an external field are also depicted.

## **Supplementary Section I:** First-principles calculations

The phase diagram of strained bulk $SrTiO_3$ is shown in Figure 5(a) of the main text. For each strain point indicated on the x axis, we tested eight different phase inputs, each consisting of a distinct combination of antiphase tilting and polar distortion (Supplementary Table I1). At many points, phases could not be relaxed. In Figure 5(a), we show the energies of all relaxed space groups at each strain. In Figure I1, we show the energy lowering of the lowest energy polar and non-polar space groups from an undistorted *P4/mmm* phase (consisting only of strain modes), as an estimate of polar transition temperature.

At 3.90 Å (bulk $SrTiO_3$ lattice parameters were found to be 3.89 Å), we observe a polar *Cc* phase ground state (consisting of antiphase tilting and polarization in-plane and out-of-plane), but only by an extremely small margin, with other polar phases *I4cm* and *Ima*2 very close in energy (shown in the zoomed inset in Figure 5(a)), and the non-polar *C2/c* phase (consisting of in-plane and out-of-plane tilting) also extremely close in energy ($\Delta E = 0.2$ meV/f.u.). Experimentally, $SrTiO_3$ is non-polar at ground state, but DFT calculations do not consider quantum fluctuations, which previous studies estimate amount to ~2 meV/f.u [9]. This value clearly exceeds the energy difference between our polar and non-polar states. Therefore, despite yielding a polar ground state, we stress that our simulations still match very well to experiment.

**Supplementary Table I1:** Space groups, the respective rotation patterns (in Glazer's notation [10]) and polar distortions of the different phases of $SrTiO_3$ examined by first-principles calculations.

| Space group | Rotation pattern | Polar distortion |
|---|---|---|
| ***P4/mmm*** | $a^0a^0a^0$ | $000$ |
| ***Cc*** | $a^-a^-c^-$ | $P_xP_yP_z$ |
| ***C2/c*** | $a^-a^-a^-$ | $000$ |
| ***I4cm*** | $a^0a^0c^-$ | $00P_z$ |
| ***I4/mcm*** | $a^0a^0c^-$ | $000$ |
| ***Ima2*** | $a^-a^-c^0$ | $P_xP_y0$ |
| ***Imma*** | $a^-a^-c^0$ | $000$ |

**Supplementary Table I2:** Unit cell tetragonality (c/a) of the different phase of strained $SrTiO_3$, in pure form and in capacitor structures computed by first-principles methods.

| | **Phase** | ***P4/mmm*** | ***I4/mcm*** | ***I4cm*** | **Stoich. capacitor (polar)** | **Non-stoich. capacitor ($NiO_2$-SrO) (non-polar)** | **Non-stoich. capacitor (NdO-$TiO_2$) (non-polar)** |
|---|---|---|---|---|---|---|---|
| **Strain** | | | | | | | |
| -1.28 % | | 1.017 | 1.024 | 1.027 | - | - | - |
| -2.56 % | | 1.034 | 1.048 | 1.057 | - | - | - |
| -2.93% | **c/a** | 1.051 | 1.056 | 1.069 | 1.072 | 1.056 | 1.056 |
| -3.85 % | | 1.053 | 1.073 | 1.091 | - | - | - |
| -5.13 % | | 1.071 | 1.101 | 1.133 | - | - | - |

In the compressive regime, we observe the rapid stabilization of the *I4cm* phase, consisting of tilt and polarization out-of-plane. The energy difference between this and the non-polar *I4/mcm* phases initially remains small for low compressive strains, but rapidly amplifies under higher strains. The energy difference quickly surpasses the scale of quantum fluctuations, suggesting at the very least the appearance of ferroelectricity at large strains and low temperatures. This is qualitatively consistent with Landau-Ginzburg-Devonshire analysis [1], though our simulations suggest the crossover of polar and tilt transition temperatures to occur at around 4-5% strain rather than the 2% strain predicted in the model. We stress that in the interfaced systems, the additional effect of the polar discontinuity limits the presence of the polar mode, while our bulk strained calculations are independent of this factor.

In the tensile regime, we observe a similar scenario, where the polar *Ima*2 phase (tilting and polarization in-plane) rapidly stabilizes as strain increases. Unlike the compressive regime however, the polar mode energies far exceed the nearest tilt-only phase (*Imma*). This matches the prediction from Landau-Ginzburg-Devonshire analysis [1].

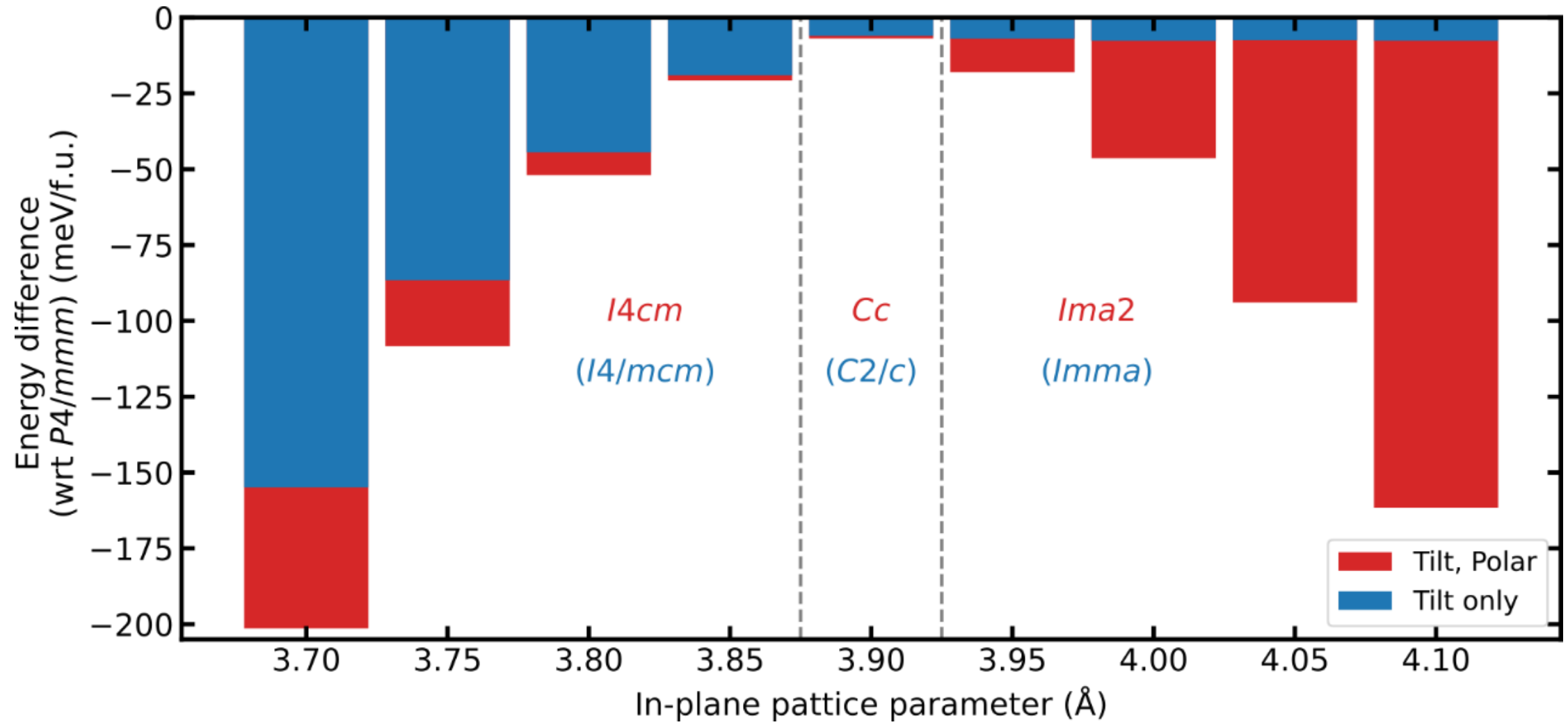

**Supplementary Figure I1:** First-principles calculations. Energy of lowest polar and non-polar (labelled in brackets) phases with respect to *P4/mmm* (001-strained $Pm\overline{3}m$).

The electronic and structural properties of the non-stoichiometric interfaces are shown in Supplementary Figures I2 and I3. The system with two SrO-$NiO_2$ interfaces (Figure I2(a)–(d)) shows the valence band maximum (VBM) of $SrTiO_3$ remaining just below the Fermi level $E_F$, while the system with two NdO-$TiO_2$ interfaces (Figure I3(a)–(d)) shows the $SrTiO_3$ VBM to lie at -1.59 eV, far from $E_F$. Such a large difference in the $SrTiO_3$ band offsets indicates that the dipoles are almost certainly not a result of intrinsic effects. Given the polar mode is not allowed to relax in $SrTiO_3$, this implies an underlying charge discontinuity of some sort is present. Further proof of this comes from the magnetic moments. Both non-stoichiometric systems are half metallic (the spin-resolved DOS only shows the spin states at the Fermi level), such that any change in the total magnetization directly corresponds to a change in the free carriers at the interfaces. The difference in total magnetization between the system with two SrO-$NiO_2$ interfaces and the system with two NdO-$TiO_2$ is 4.055$\mu_B$. If we were to assume $NdNiO_3$ and $SrTiO_3$ resembled their nominal charges, such that they were III/III and II/IV perovskites respectively, the polar discontinuity would be 0.5*e* per interfacial f.u.. In our systems, given there are two interfacial formula units, and two interfaces each, there would be 2*e* bound charge per system. Therefore, the difference in hypothetical bound charge between the two interfaced systems would be 2*e* + 2*e* = 4*e*. This is a very close match to the difference in total magnetization. This suggests that, for the purpose of modelling the electrostatics, the interface between $NdNiO_3$ and $SrTiO_3$ could be viewed as a sum of bound and free charges each of 0.5e per interfacial f.u. (of equal and opposite sign, such that the net interface charge density is zero, as expected for a metal) extremely similar to that between $LaAlO_3$ and $SrTiO_3$ [11,12] (two insulating III/III and II/IV perovskites respectively).

Overall, this provides strong evidence that in the stoichiometric system (main text Figure 5(b)–(e)) with inversion symmetry broken, the large field is caused by this same underlying charge discontinuity. In the stoichiometric system, this field is partially screened by the $SrTiO_3$ polar mode, which becomes fixed in one direction. This fully supports what is observed experimentally, where no $SrTiO_3$ polar reversal is possible, while the $SrTiO_3$ inter-layer distance is observed to be significantly amplified, which typically indicates the presence of a polar distortion.

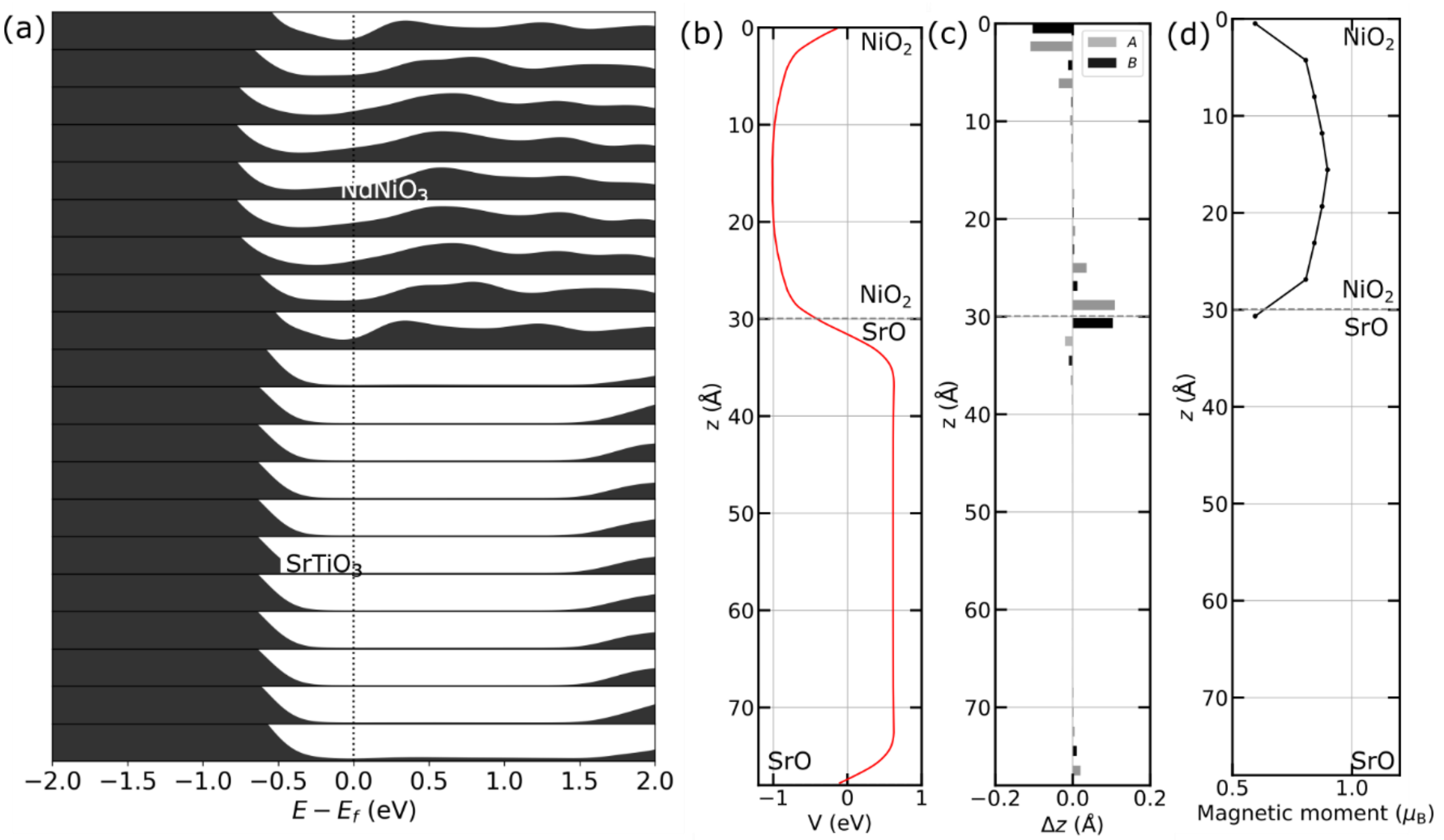

**Supplementary Figure I2:** Electronic and structural characteristics of a non-stoichiometric $NdNiO_3$-$SrTiO_3$ system with two $NiO_2$-SrO interfaces as a function of z position (direction along $\hat{c}_{001}$): (a) layer projected density of states (PDOS) near $E_F$, where each layer has chemical formula $2ABO_3$, and the y-axis of each layer plot ranges from 0 to 3 states/eV, (b) macroscopically averaged potential (averaged over $SrTiO_3$ interlayer distance (= 4.0 Å)), (c) A- and B-site polar displacements and (d) layer magnetic moments along $\hat{c}_{001}$.

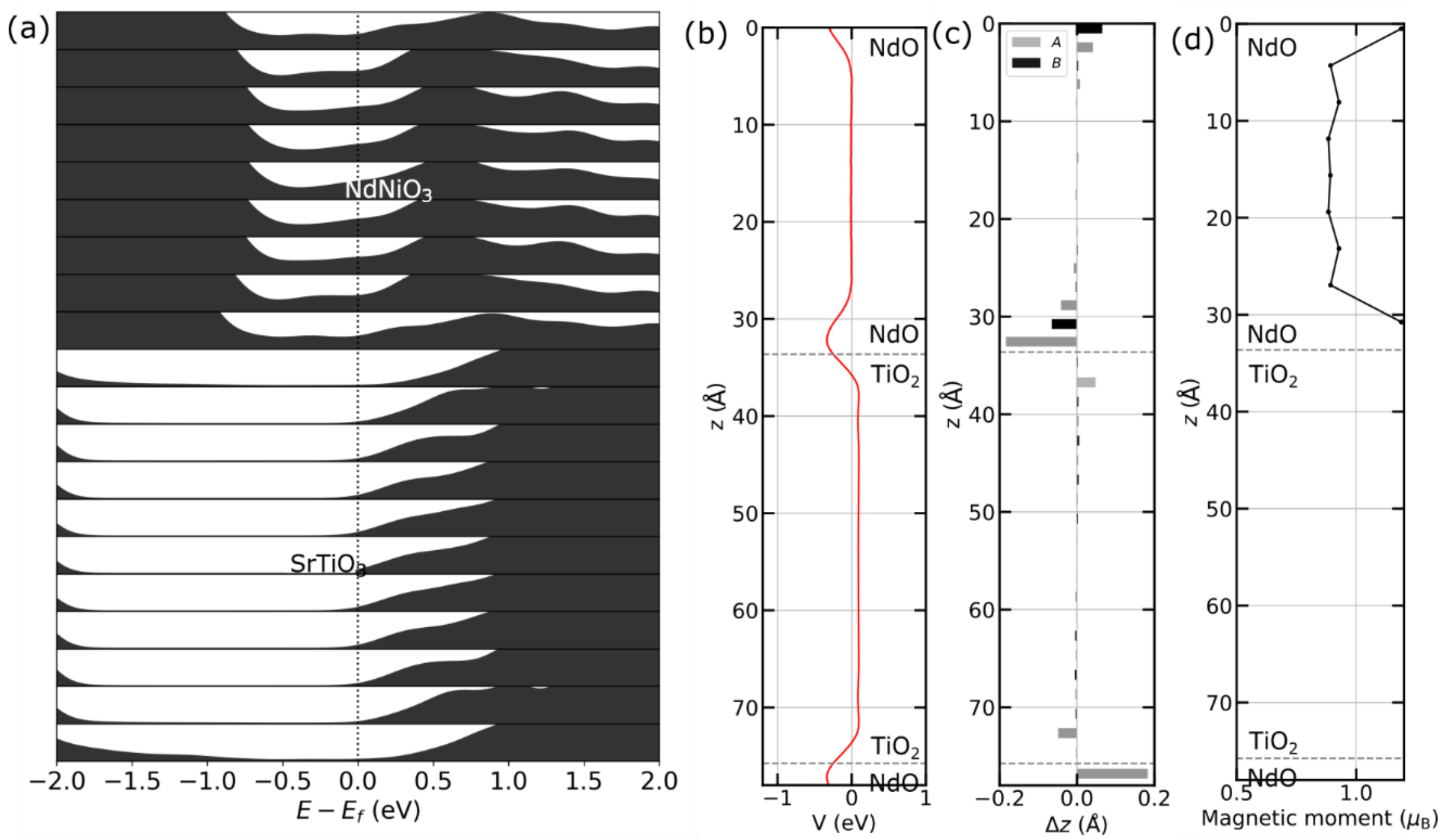

**Supplementary Figure I3:** Electronic and structural characteristics of a non-stoichiometric $NdNiO_3$-$SrTiO_3$ system with two NdO-$TiO_2$ interfaces as a function of z position (direction along $\hat{c}_{001}$): (a) layer projected density of states (PDOS) near $E_F$, where each layer has chemical formula $2ABO_3$, and the y-axis of each layer plot ranges from 0 to 3 states/eV, (b) macroscopically averaged potential (averaged over $SrTiO_3$ interlayer distance (= 4.0 Å)), (c) A- and B-site polar displacements and (d) layer magnetic moments along $\hat{c}_{001}$.